\documentclass[12pt]{article}

\title{Optimal consumption and investment
 for  markets with  random coefficients.
\thanks{The paper is  supported by the RFBR-Grant 09-01-00172-a.}
}

\author{
Belkacem Berdjane\thanks{
Laboratoire L2CSP, de l'universit\'e de Tizi-Ouzou (Alg\'erie) \& Laboratoire LMRS, de l'universit\'e de Rouen (France); \, email: berdjane\_b@yahoo.fr}
 and
 Serguei Pergamenshchikov\thanks{
 Laboratoire de Math\'ematiques Raphael Salem,
 Avenue de l'Universit\'e, BP. 12,
  Universit\'e de Rouen,
   F76801, Saint Etienne du Rouvray, Cedex France
and Department of Mathematics and Mechanics,Tomsk State University,
Lenin str. 36, 634041 Tomsk, Russia, e-mail:
Serge.Pergamenchtchikov@univ-rouen.fr } }

\usepackage{srcltx}
\usepackage{epsfig}
\usepackage{graphicx}

\usepackage{amssymb}
\usepackage{amsfonts}
\usepackage{amsmath}
\usepackage{amsthm}
\usepackage{enumerate}
\usepackage{multicol}

\newtheorem{theorem}{Theorem}[section]
\newtheorem{proposition}[theorem]{Proposition}
\newtheorem{lemma}[theorem]{Lemma}
\newtheorem{definition}[theorem]{Definition}
\newtheorem{remark}{Remark}[section]
\newtheorem{corollary}[theorem]{Corollary}

\newcommand\cE{{\cal E}}
\newcommand\cH{{\cal H}}
\newcommand\cG{{\cal G}}
\newcommand\cF{{\cal F}}

\newcommand\cL{{\cal L}}
\newcommand\cB{{\cal B}}
\newcommand\cK{{\cal K}}

\newcommand\cX{{\cal X}}

\newcommand\cT{{\cal T}}

\newcommand\cV{{\cal V}}

\newcommand\0{{\bf 0}}

\def\bbr{{\mathbb R}}

\def\bbma{{\mathbb M}}

\def\text#1{\hbox{#1}}
\def\proof{{\noindent \bf Proof. }}
\def\endproof{\mbox{\ $\qed$}}

\def\a{{\bf a}}
\def\B{{\bf B}}
\def\A{{\bf A}}
\def\b{{\bf b}}
\def\q{{\bf q}}
\def\r{{\bf r}}
\def\p{{\bf p}}

\def\f{{\bf f}}
\def\h{{\bf h}}

\def\g{{\bf g}}
\def\E{{\bf E}}
\def\P{{\bf P}}

\def\C{{\bf C}}
\def\D{{\bf D}}
\def\G{{\bf G}}
\def\H{{\bf H}}
\def\M{{\bf M}}
\def\U{{\bf U}}
\def\L{{\bf L}}

\newcommand{\wh}{\widehat}
\newcommand{\wt}{\widetilde}

\def\Chi{{\bf 1}}
\def\tr{\mathrm{tr}}

\def\d{\mathrm{d}}
\def\build #1_#2{\mathrel{\mathop{\kern 0pt #1}\limits_{#2}}}
\newcommand{\zs}[1]{{\mathchoice{#1}{#1}{\lower.25ex\hbox{$\scriptstyle#1$}}
{\lower0.25ex\hbox{$\scriptscriptstyle#1$}}}}

\numberwithin{equation}{section}
\begin{document}
\maketitle

\abstract{
We consider an  optimal 
investment and consumption problem for  a
Black-Scholes  financial market 
with stochastic coefficients driven by a diffusion process. 
We assume that an agent makes consumption and investment decisions based
on CRRA utility functions. The dynamical programming approach
leads to an investigation of 
the  Hamilton Jacobi Bellman (HJB) equation which 
is a highly non linear partial differential equation (PDE) of the second oder. 
By using the Feynman - Kac representation we prove uniqueness
and smoothness of the solution. Moreover, we 
study the optimal convergence rate of
the iterative numerical schemes
for both the value function and the optimal portfolio. 
 We show, that in this case, the optimal convergence rate 
is super geometrical, i.e. is more rapid than any geometrical one.
 We apply our results to a stochastic volatility financial market.
}


\vspace*{2mm}


\section{Introduction}\label{sec:In}

One of the principal questions in mathematical finance is the optimal 
consumption-investment
problem for continuous time market models.  
This paper deals with an investment problem aiming at optimal consumption during a fixed
investment interval $[0,T]$ in addition to an optimal terminal wealth at maturity $T$.
Such problems are of prime interest for the institutional investor, selling asset funds
to their customers, who are entitled to certain payment during the duration of an
investment contract and expect a high return at maturity. The classical approach to
this problem goes back to Merton \cite{Merton1971} and involves utility functions,
 more precisely, the expected utility serves as the functional which has to be optimized.
By applying results from the stochastic control, explicit solutions have been obtained for
financial markets with nonrandom coefficients
 (see, e.g. \cite{KaratzasShreve1998}, \cite{Korn1997} and 
references therein). Since then, there has been a growing interest in consumption and 
investment problems and the Merton problem has been extended in many directions. 
One of the generalisations considers financial models with random coefficients such as 
stochastic volatility markets (see, e.g., \cite{FouquePapanicoloauSircar2000}). 
 In this paper for
the CRRA (Constant Relative Risk Aversion) utility functions
we consider 
the optimal consumption-investment problem 
for a Black-Scholes type  model with
 coefficients depending on a diffusion process which is 
 referred as the external stochastic factor.
The pure 
investment problem for such models is considered in 
\cite{Zariphopoulou2001} and
\cite{Pham2002}. In these papers the authors use 
the dynamic programming 
approach and they show that the nonlinear HJB equation can be transformed 
in a
quasilinear PDE.
The similar approach has been used in \cite{KraftSteffensen2006} 
for optimal consumption-investment problems with the default risk
for financial markets with non random coefficients.
 Furthermore, in  \cite{FlemingHernandezHernandez2003},
by making use of the Girsanov
 measure transformation
the authors study a pure optimal consumption problem 
for stochastic volatility financial markets.
 In 
\cite{CastanedaLeyvaHernandezHernandez2005} and \cite{HernandezHernandezShied2006}
 the authors use  dual methods.
Usually, the classical existence and uniqueness theorem
for the HJB equation is shown by the linear PDE methods 
 (see, for example, chapter VI.6 and appendix E in \cite{FlemingRishel1975}).
In this paper we use the approach proposed in \cite{DelongKluppelberg2008}
for the optimal consumption-investment problem for financial markets
with  random coefficients depending on pure jumps processes.
Unfortunately, we can not apply directly this method for our case since in 
\cite{DelongKluppelberg2008}
the HJB equation is the integro-differential equation of the first oder.
In our case it is a highly non linear PDE of the second oder. Similarly to
\cite{DelongKluppelberg2008} we study the HJB equation  through
the Feynman - Kac representation. We introduce a special metric space
in which the Feynman - Kac mapping is contracted. Taking this into account we show 
the fixed-point theorem for this mapping and we show that the fixed-point solution
is the classical unique solution for HJB equation in our case. Moreover,
using the verification theorem  we provide the explicite expressions
for the optimal investment and consumption which depend on the HJB solution. 
Therefore, to calculate the optimal strategies one needs to study  numerical schemes
for HJB equation. To this end we find an explicite upper bound for the
approximation accuracy. Then, we minimize this bound and we get the optimal
 convergence rate for both the value function
 and the optimal financial strategies. It turns out that in this case 
 this rate is super geometrical. 


The rest
of the paper is organized as follows. In section~\ref{sec:Mm}
we introduce the financial market, we state the main conditions on the
market parameters and we write the HJB equation. 
In section~\ref{sec:Mr} we state the
main results of the paper.
Section~\ref{sec:Sv}
presents a stochastic volatility model
as an example of applications of our results.
In Section~\ref{sec:PrL} we study the properties
of the Feynman - Kac mapping.
In Section~\ref{sec:Vt}
the corresponding verification theorem is stated.
 The proofs of the main results are given in
Section \ref{sec:Pr}. In Section~\ref{sec:Nu}
we consider a numerical example.
In Appendix some auxiliary results
are given.

\section{Market model}\label{sec:Mm}

Let $(\Omega, \cF_\zs{T}, (\cF_\zs{t})_\zs{0\le t\le T},\P)$
be a standard filtered probability space
with two standard independent $(\cF_\zs{t})_\zs{0\le t\le T}$ adapted
Wiener processes $(W_\zs{t})_\zs{0\le t\le T}$ and
$(V_\zs{t})_\zs{0\le t\le T}$ taking their values in $\bbr^{d}$ and
$\bbr^{m}$ respectively, i.e.
$$
W_\zs{t}=(W^{1}_\zs{t},\ldots,W^{d}_\zs{t})'
\quad
\mbox{and}
\quad
V_\zs{t}=(V^{1}_\zs{t},\ldots,V^{m}_\zs{t})'
\,.
$$
The prime $'$ denotes the transposition.
Our financial market consists of one
{\em riskless bond}  $(S_\zs{0}(t))_\zs{0\le t\le T}$ and
$d$  {\em risky stocks}
$(S_\zs{i}(t))_\zs{0\le t\le T}$ governed by the following equations:
\begin{equation}\label{sec:Mm.1}
\left\{\begin{array}{ll}
\d S_\zs{0}(t)&=r(t,Y_\zs{t})\,S_\zs{0}(t)\d t\,,\\[5mm]
\d S_\zs{i}(t)&=S_\zs{i}(t)\mu_\zs{i}(t,Y_\zs{t})\d t+S_\zs{i}(t)\,
\sum^d_\zs{j=1}\,\sigma_\zs{ij}(t,Y_\zs{t})\,\d W^{j}_\zs{t}\,,
\end{array}\right.
\end{equation}
with $S_\zs{0}(0)=1$ and $S_\zs{i}(0)=s_\zs{i}$ for $1\le i\le d$.
In this model
$r(t,y)\in\bbr_\zs{+}$ is the {\em riskless interest rate},
$\mu(t,y)=(\mu_\zs{1}(t,y),\ldots,\mu_\zs{d}(t,y))'$ is the vector of
{\em stock-appreciation rates} and
$\sigma(t,y)=(\sigma_\zs{ij}(t,y))_\zs{1\le i,j\le d}$ is
the  matrix of {\em stock-volatilities}.
For all  $y\in\bbr^{m}$
the coefficients $r(\cdot,y)$, $\mu(\cdot,y)\in\bbr^{d}$ and
$\sigma(\cdot,y)\in\bbma_\zs{d}$ are
nonrandom c\`adl\`ag functions. Here $\bbma_\zs{d}$ denotes  the set of quadratic matrix
of order $d$. 
Moreover, in just the same way as in \cite{Zariphopoulou2001} 
we assume, that the stochastic factor $Y$ valued in $\bbr^{m}$ 
has a dynamics governed by the
following stochastic differential equation:
\begin{equation}\label{sec:Mm.2}
\d Y_\zs{t}=F(t,Y_{t})\,\d t+\beta\d\U_{t}\,,
\end{equation}
where $F$ is a  $[0,T]\times \bbr^{m}\to \bbr^{m}$  nonrandom function 
and $\beta$ is fixed positive parameter. The process $U$
is the standard Brownian motion defined as 
\begin{equation}\label{sec:Mm.2-1}
\U_{t}=\rho V_\zs{t}+\sqrt{1-\rho^{2}}\sigma_\zs{*} W_\zs{t}\,,
\end{equation}
where $0\le \rho\le 1$ and
$\sigma_\zs{*}$ is a fixed
 $m\times d$ matrix for which
$\sigma_\zs{*}\sigma'_\zs{*}=I_\zs{m}$.
Here $I_\zs{m}$ is the identity matrix of order $m$.

Moreover, we set $\cK=[0,T]\times \bbr^{m}$ and we note,  
that for the model \eqref{sec:Mm.1} the risk premium is the
$\cK\to\bbr^{d}$ function defined as
\begin{equation}\label{sec:Mm.3}
\theta(t,y)=\sigma^{-1}(t,y)(\mu(t,y)-r(t,y)\,\Chi_\zs{d})\,,
\end{equation}
where
$\Chi_\zs{d}=(1,\ldots,1)'\in\bbr^d$.
Similarly to \cite{KluppelbergPergamenchtchikov2009} we consider the fractional portfolio process
$$
\varphi(t)=(\varphi_\zs{1}(t),\ldots,\varphi_\zs{d}(t))'\in\bbr^d\,,
$$
i.e. $\varphi_\zs{i}(t)$ represent the fraction of the wealth process $X_\zs{t}$ invested in
the $i$-th stock at the time $t$. The fractions for consumptions 
we denote by $c=(c_\zs{t})_\zs{0\le t\le T}$. In this case  the wealth process
satisfies the following stochastic equation
\begin{equation}\label{sec:Mm.5}
 \d X_\zs{t}=X_\zs{t}(r(t,Y_\zs{t})+\pi'_\zs{t}
\theta(t,Y_\zs{t})-c_\zs{t})
\d t+X_\zs{t}\pi'_\zs{t}\d W_\zs{t}\,,
\end{equation}
where $\pi_\zs{t}= \sigma(t,Y_\zs{t})\varphi_\zs{t}$ and the initial endowment
$X_\zs{0}=x$.
 Now we describe the set of all admissible
strategies. A portfolio control (financial strategy)
$\vartheta=(\vartheta_\zs{t})_\zs{t\ge 0}=((\pi_\zs{t},c_\zs{t}))_\zs{t\ge 0}$ 
is  said to be {\em admissible} if it is $(\cF_\zs{t})_\zs{0\le t\le T}$ - progressively measurable with values in $\bbr^d\times [0,\infty)$, such that
\begin{equation}\label{sec:Mm.6}
\|\pi\|_\zs{T}
:=\int^{T}_\zs{0}
|\pi_\zs{t}|^2\d t
<\infty
\quad\mbox{and}\quad
\int^{T}_\zs{0}\,c_\zs{t}\d t
<\infty
\quad\mbox{a.s.}
\end{equation}
and the equation \eqref{sec:Mm.5} has a unique strong a.s. positive continuous solution
$(X^{\vartheta}_\zs{t})_\zs{0\le t\le T}$ on $[0\,,\,T]$.
We denote the set 
of  {\em admissible portfolios controls}
 by $\cV$.

In this paper we consider an agent using the CRRA
utility function $x^{\gamma}$ for $0<\gamma<1$.
The goal is to maximize the expected utilities from the consumption
 on the time interval $[0,T]$ and from the  terminal wealth.
Then for any $x\in\bbr^{d}$, $y\in\bbr^{m}$ and $\vartheta\in\cV$ the
 value function of agent is 
$$
J(x,y,\vartheta):=\E_{x,y}\,\left(\int^T_\zs{0}
c_\zs{t}^{\gamma}\,(X^{\vartheta}_\zs{t})^{\gamma}\d t\,+\,(X^{\vartheta}_\zs{T})^{\gamma}\right)\,,
$$
were $\E_\zs{x,y}$ is the conditional expectation
for  $X^{\vartheta}_\zs{0}=x$ and $Y_\zs{0}=y$. 
Our goal is to maximize this function, i.e.
\begin{equation}\label{sec:Mm.7}
\sup_\zs{\vartheta\in\cV}\,J(x,y,\vartheta)\,.
\end{equation}


\begin{remark}\label{Re.sec:Mm.1}
Note, that
 \eqref{sec:Mm.7} is the classical
 Merton optimization problem for the market \eqref{sec:Mm.1},
 (see, e.g. \cite{Merton1971}, \cite{DelongKluppelberg2008}).
Pure investment cases of this problem are studied in 
 \cite{Zariphopoulou2001} and \cite{Pham2002}.
\end{remark}

\subsection{Conditions}

\noindent To list  the conditions for the model \eqref{sec:Mm.1}
we set
\begin{equation}\label{sec:Mm.8}
\alpha (t,y)=F(t,y)+
\,
\beta_\zs{*}\,
\sigma_\zs{*}\theta (t,y)\,,
\end{equation}
where 
$\beta_\zs{*}=\gamma\sqrt{1-\rho^{2}}\beta/(1-\gamma)$.
Moreover, we denote by $\C^{i,k}(\cK)$ the space of the functions 
$f(t,y_\zs{1},\ldots,y_\zs{m})$ which are  $i$ times
differentiable with respect to $t\in [0,T]$ and $j$ times differentiable
 with respect to $(y_\zs{j})_\zs{1\le j\le m}$. 

\medskip

\noindent We assume that the market parameters satisfy the following conditions: \\[3mm]
\noindent $\A_\zs{1})$
{\sl The functions
$r:\cK \rightarrow \bbr_\zs{+}$,
$\mu:\cK\rightarrow \bbr_\zs{+}^{d}$
 and
$\sigma :\cK \rightarrow \bbma_\zs{d}$ belong to
$\C^{1,1}(\cK)$ and have bounded derivatives. 
Moreover, for  all $(t,y)\in\cK$
 the matrix $\sigma(t,y)$ is non-degenerated
 and
$$
\sup_\zs{(t,y)\in \cK}\,
\left(
|r(t,y)|
+
|\mu(t,y)|
+
|\sigma^{-1} (t,y)|
\right)<\infty\,.
$$
}

\medskip

\noindent $\A_\zs{2})$
 {\sl The $\cK\to\bbr^{m}$ function $F(\cdot,\cdot)$  belongs to
$\C^{1,1}(\cK)$  and its  partial derivatives are  bounded.}

\noindent This condition provides the existance 
of a unique strong solution for the equation  \eqref{sec:Mm.2}.
On the interval $[t,T]$ we denote this solution  by

$Y^{t,y}=(Y^{t,y}_\zs{v})_\zs{t\le v\le T}$ with
$Y^{t,y}_\zs{t}=y$.

\medskip

\noindent $\A_\zs{3})$
 {\sl There exist $[0,T]\times\bbr\to\bbr$ continuously differentiable
  functions 
$(\a_\zs{i,j}(\cdot,\cdot))_\zs{1\le i,j\le m}$ such that 
for any $0\le t\le T$ and for any 
$x=(x_\zs{1},\ldots,x_\zs{m})'$ from $\bbr^{m}$
 the $i$th component $[\alpha(\cdot,\cdot)]_\zs{i}$ of the vector
 $\alpha(t,x)$ 
has the form
\begin{equation}\label{sec:Mm.9}
[\alpha(t,x)]_\zs{i}=\sum^{m}_\zs{l=1}\,\a_\zs{i,l}(t,x_\zs{j})
\end{equation}
and, moreover, 
$$
\alpha_\zs{*}=
\max_\zs{i,j}\,
\max_\zs{t,z}\,
\max
\left(
|\a_\zs{i,j}(t,z)|
\,,\,
\left|\frac{\partial \a_\zs{i,j}(t,z)}{\partial t}\right|
\,,\,
\left|
\frac{\partial \a_\zs{i,j}(t,z)}{\partial z}
\right|
\right)
<\infty\,.
$$
}
\medskip

\noindent

 In the sequel, for any vector $x$ from $\bbr^{n}$
we denote the $i$th component by $[x]_\zs{i}$.

\begin{remark}\label{Re.sec:Mm.1}
 In Section~\ref{sec:Sv} we check the condition $\A_\zs{3}$)
 for a two asset stochastic volatility
financial market.
\end{remark}

\medskip
\medskip

\subsection{The HJB equation}\label{sec:HJB}

Now we introduce the HJB equation for the problem 
\eqref{sec:Mm.7}.
To this end, for any differentiable $\cK\to\bbr$
function $f$ we  denote by  $\D_\zs{y} f(t,y)$  its 
 gradient with respect to $y\in\bbr^{m}$, i.e. 
$$
\D_\zs{y} f(t,y)=
\left(
\frac{\partial  }{\partial y_\zs{1}}f(t,y),\ldots,
\frac{\partial }{\partial y_\zs{m}}f(t,y)
\right)'\,.
$$
Let now $z(t,\varsigma)$ be any $[0,T]\times\bbr\times\bbr^{m}\to\bbr$
two times differentiable  function. 
Here $\varsigma=(x,y)'$, $x\in\bbr$ and $y\in\bbr^{m}$.
We set 
$$
\D_\zs{\varsigma}z(t,\varsigma)=
\left(
z_\zs{x}(t,\varsigma)\,,\,
\D_\zs{y}z(t,\varsigma)
\right)'
$$
and
$$
\D_\zs{\varsigma,\varsigma}z(t,\varsigma)
=\left(
\begin{array}{cc}
\dfrac{\partial^{2} z(t,\varsigma)}{\partial^{2}x}\,;& (\D_\zs{x,y}z(t,\varsigma))' \\[4mm]
\D_\zs{x,y}z(t,\varsigma)\,;&\,\D_\zs{y,y}z(t,\varsigma)
\end{array}
\right)\,,
$$
where
$$
\D_\zs{x,y}z(t,\varsigma)=
\left(
\frac{\partial^{2} z(t,\varsigma) }{\partial x \partial y_\zs{1}}\,,\ldots,
\frac{\partial^{2} z(t,\varsigma) }{\partial x\partial y_\zs{m}}
\right)'
$$
and
$$
\D_\zs{y,y}z(t,\varsigma)=\left(
\frac{\partial^{2}  z(t,\varsigma)}{\partial y_\zs{j}\partial y_\zs{i}}
\right)_\zs{1\le i,j\le m}\,.
$$
Let now  $\q\in\bbr^{m+1}$ and $\M\in\bbma_\zs{m+1}$ 
be fixed parameters of the following form
\begin{equation}\label{sec:HJB.0}
 \q=(\q_\zs{1},\wt{\q})'
\quad\mbox{and}\quad
\M
=\left(
\begin{array}{cc}
\mu\,;& \wt{\mu}' \\[2mm]
\wt{\mu}\,;&\,\M_\zs{0}
\end{array}
\right)\,,
\end{equation}
where $\q_\zs{1},\mu\in\bbr$,
$\wt{\q},\wt{\mu}\in\bbr^{m}$
and $\M_\zs{0}\in\bbma_\zs{m}$.
For these parameters with $\q_\zs{1}>0$ we 
define the Hamilton function as
\begin{align}\nonumber
H(t,\varsigma,\q,\M)
&=
x\,r(t,y)\q_\zs{1}+(F(t,y))^{'}\wt{\q}
+\frac{1}{\gamma_\zs{1}}
\left(\frac{\gamma}{\q_\zs{1}}\right)^{\gamma_\zs{1}-1}
\\[2mm]\label{sec:HJB.1}
&+ \frac{|\theta(t,y)\q_\zs{1}
+
\beta\sqrt{1-\rho^{2}}\sigma'_\zs{*}\wt{\mu}|^{2}}{2|\mu|}
+\frac{\beta^{2}}{2}\tr \M_\zs{0}\,,
\end{align}
where $\gamma_\zs{1}=(1-\gamma)^{-1}$.
 In this case the HJB equation is
\begin{equation}\label{sec:HJB.2}
 \left\{
\begin{array}{rl}
&z_\zs{t}(t,\varsigma)+
H(t,\varsigma,\D_\zs{\varsigma}z(t,\varsigma),
\D_\zs{\varsigma,\varsigma}z(t,\varsigma))=0\,, \\[3mm]
&z(T,\varsigma)=x^{\gamma}\,.
\end{array}
\right.
\end{equation}
To study this equation  we make use of the distortion 
power transformation introduced in \cite{Zariphopoulou2001}, i.e. 
for a positive function $h$ we represent $z(t,y)$ as
\begin{equation}\label{sec:HJB.3}
z(t,\varsigma)=x^{\gamma}h^{\varepsilon}(t,y)
\,,\quad
\varepsilon =\frac{1-\gamma }{1-\rho^{2}\gamma }\,.
\end{equation}
It is easy to deduce that the function $h$ satisfies the following quasi-linear PDE: 
\begin{equation}\label{sec:HJB.4}
\left\{
\begin{array}{rl}
h_\zs{t}(t,y)&+
Q(t,y) h(t,y)
+
(\alpha(t,y))'\D_\zs{y}\,h(t,y)\\[5mm]
&
+\dfrac{\beta^{2}}{2}\tr \D_\zs{y,y}h(t,y)
+\dfrac{1}{q_\zs{*}}
\left(\dfrac{1}{h(t,y)}\right)^{q_\zs{*}-1}
=0\,;
\\[5mm]
h(T,y)&=1\,,
\end{array}
\right.
\end{equation}
where
\begin{equation}\label{sec:Mr.1}
Q(t,y)=
\frac{\gamma(1-\rho^{2}\gamma)}{1-\gamma}
\,
\left( r(t,y)+
\frac{|\theta(t,y)|^{2}}{2\left(1-\gamma\right)}\right)\,.
\end{equation}
Note that, by the condition $\A_\zs{1})$ this function and its $y$-gradient are bounded.
Therefore, we can set
\begin{equation}\label{sec:Mr.2-1}
Q_\zs{*}=\sup_\zs{(t,y)\in \cK}\,Q(t,y)
\quad\mbox{and}\quad
D_\zs{*}=
\sup_\zs{(t,y)\in \cK}\,
|\D_\zs{y} Q(t,y)|\,.
\end{equation}

Our  goal is to study the equation \eqref{sec:HJB.4}. 
By making use of the probabilistic representation for 
the linear PDE (the Feynman-Kac formula)
 we  show in Proposition~\ref{Pr.sec:Prl.5},  that the solution of this equation is 
the fixed-point solution for a special mapping of the integral type
which will be introduced in the next section.

\medskip
\medskip
\section{Main results}\label{sec:Mr}

First, to study the equation  \eqref{sec:HJB.4}
 we  introduce a special  functional space. Let $\cX$ be the set of
$\cK\to [1,\infty)$ functions
from $\C^{0,1}\left(\cK\right)$
 such that
\begin{equation}\label{sec:Mr.1-1}
\|f\|_\zs{0,1}=
\sup_\zs{(t,y)\in\cK}\,
\left(
|f(t,y)|+
|\D_\zs{x}f(t,y)|
\right)
\,
\le \r^{*}
\,,
\end{equation}
where  
\begin{equation}\label{sec:Fr.4}
\r^{*}=
\r_\zs{0}
+
m\left(
\frac{H^{*}}{q_\zs{*}}(2\sqrt{T}+T)
+D_\zs{*}T
e^{(\alpha_\zs{*}+Q_\zs{*})T}
\right)\,,
\end{equation}
$\r_\zs{0}=
e^{Q_\zs{*}T}
+
(e^{Q_\zs{*}T}-1)/Q_\zs{*}q_\zs{*}$. The parameter $H^{*}$ will be 
defined below in Section~\ref{sec:Fr}.
To define a metrics in $\cX$ we set
\begin{equation}\label{sec:Mr.2}
\varkappa=Q_\zs{*}+\zeta+1+m\L_\zs{*}
\quad\mbox{and}\quad
\L_\zs{*}= (1+\r^{*}q_\zs{*}+
T D_\zs{*}
)e^{\alpha_\zs{*}T}
\,,
\end{equation}
where the positive parameter $\zeta$ will be specified
later,
$q_\zs{*}=(1-\rho^{2}\gamma)^{-1}$.
Now 
 for any $f$ and  $g$ from $\cX$ we define the metrics as 
\begin{equation}\label{sec:Mr.3}
\varrho_\zs{*}(f,g)= \|f-g\|_\zs{*}
\,,
\end{equation}
where
$$
\|f\|_\zs{*} =\sup_\zs{(t,y)\in \cK}
\,
e^{-\varkappa(T-t)}
\,
\left(
|f(t,y)|+
|\D_\zs{x}f(t,y)|
\right)
\,.
$$

\noindent 
Moreover,
to write the Feynman-Kac representation for the equation \eqref{sec:HJB.4}
we need to use the stochastic process $(\eta_\zs{v})_\zs{0\le v\le T}$ 
governed by 
the following stochastic differential equation
\begin{equation}\label{sec:Mr.4}
\d\eta_\zs{t}=
\alpha (t,\eta_\zs{t})\d t+\beta\,d\U_\zs{t}\,.
\end{equation}
\noindent
It should be noted that the condition $\A_\zs{3})$ provides
the existence and the
uniqueness  of a strong solution on any time interval 
$[t,T]$ and initial condition $y$ from $\bbr^{m}$. 
We denote this solution by $(\eta^{t,y}_\zs{v})_\zs{t\le v\le T}$. 
 Using this process  we define  the  $\cX \rightarrow \cX$ Feynman-Kac 
mapping $\cL $:
\begin{equation}\label{sec:Mr.5}
\cL_\zs{f}(t,y)= \E\,\cG(t,T,y)
+\frac{1}{q_\zs{*}}
\int_{t}^{T}
\cH_\zs{f}(t,s,y)
\,\d s\,,
\end{equation}
where $\cG(t,s,y)= \exp \left( \int_{t}^{s} Q(u,\eta_{u}^{t,y})\d u\right)$ and
\begin{equation}\label{sec:Mr.6}
\cH_\zs{f}(t,s,y)=\E
\left(f(s,\eta^{t,y}_{s})\right)^{1-q_\zs{*}}
\cG(t,s,y)\,.
\end{equation}
To solve the  equation \eqref{sec:HJB.4} we need to
find the fixed-point solution for the mapping $\cL$ in
$\cX$, i.e.
\begin{equation}\label{sec:Mr.7}
\cL_\zs{h}=h\,.
\end{equation}
To this end we construct
the following  iterated scheme. We set
$h_\zs{0}\equiv 1$ and
\begin{equation}\label{sec:Mr.8}
h_\zs{n}(t,y)=\cL_\zs{h_\zs{n-1}}(t,y)
\quad\mbox{for}\quad n\ge 1\,.
\end{equation}

\noindent First, we study the behavior of the deviation
$$
\Delta_\zs{n}=h(t,y)-h_\zs{n}(t,y)\,.
$$

\noindent In the following theorem  we show that the contraction coefficient
for the operator \eqref{sec:Mr.5}
is given by
\begin{equation}\label{sec:Mr.9}
\lambda=\frac{1+m\L_\zs{*}}{\zeta + 1+m\L_\zs{*}}
\end{equation}
and  an appropriate choice of $\zeta$ gives the super-geometrical convergence rate
for the sequence $(h_\zs{n})_\zs{n\ge 1}$.

\begin{theorem}\label{Th.sec:Mr.1}
Under the conditions 
$\A_\zs{1}$)--$\A_\zs{3}$) the equation \eqref{sec:Mr.7} has a unique solution $h$ in $\cX$
such that for any $n\ge 1$ and $\zeta>0$
\begin{equation}\label{sec:Mr.10}
\sup_\zs{(t,y)\in \cK}
\left(
|\Delta_\zs{n}(t,y)|
+
\left|
\D_\zs{y} \Delta_\zs{n}(t,y)
\right|
\right)
\le \B^{*}\,\lambda^{n}\,,
\end{equation}
where
$\B^{*}=e^{\varkappa T}\,(1+\r^{*})/(1-\lambda)$
and $\varkappa$ is given 
in \eqref{sec:Mr.2}.
\end{theorem}
\noindent
The proof of this theorem is given in Section~\ref{sec:Pr}. 

Now we can minimize the upper bound \eqref{sec:Mr.10} over $\zeta>0$. Indeed,
setting 
$\wt{\zeta}=\zeta/(1+m\L_\zs{*})$ and $\wt{T}=(1+m\L_\zs{*}) T$,
 we obtain
$$
\B^{*}\,\lambda^{n}=\C^{*}\,
\exp\{\g_\zs{n}(\wt{\zeta})\}\,,
$$
where 
$\C^{*}=(1+\r^{*})e^{(Q_\zs{*}+1+m\L_\zs{*})T}$
and
$$
\g_\zs{n}(x)=x\wt{T}-\ln x-(n-1)\ln(1+x)\,.
$$
Now we minimize this function over $x>0$, i.e.
$$
\min_\zs{x>0} \g_\zs{n}(x)
=
x^{*}_\zs{n}\wt{T}-
\ln x^{*}_\zs{n}-(n-1)\ln(1+x^{*}_\zs{n})
\,,
$$
where
$$
x^{*}_\zs{n}=\frac{\sqrt{(\wt{T}-n)^2+4\wt{T}}+n-\wt{T}}{2\wt{T}}\,.
$$
Therefore, for 
$$
\zeta=\zeta^{*}_\zs{n}=(1+m\L_\zs{*})x^{*}_\zs{n}
$$
 we obtain  the optimal upper bound \eqref{sec:Mr.10}.
\begin{corollary}\label{Co.sec:Mr.1}
Under the conditions 
$\A_\zs{1}$)--$\A_\zs{3}$) the equation \eqref{sec:Mr.7} has a unique solution $h$ in $\cX$
such that for any $n\ge 1$
\begin{equation}\label{sec:Mr.11}
\sup_\zs{(t,y)\in \cK}
\left(
|\Delta_\zs{n}(t,y)|
+
\left|
\D_\zs{y} \Delta_\zs{n}(t,y)
\right|
\right)
\le \U^{*}_\zs{n}\,,
\end{equation}
where 
$\U^{*}_\zs{n}=\C^{*}\,\exp\{\g^{*}_\zs{n}\}$.
\end{corollary}

\begin{remark}\label{Re.sec:Mr.1}
One can check directly that for some $\delta>0$
$$
\U^{*}_\zs{n}=\mbox{O}(n^{-\delta n})\quad\mbox{as}\quad n\to\infty\,.
$$
This means that the convergence rate is more rapid than any 
geometrical one, i.e. it is  super geometrical.
\end{remark}

\medskip

\begin{theorem}\label{Th.sec:Mr.2}
Assume, that the conditions 
$\A_\zs{1}$)--$\A_\zs{3}$) hold.
Then the optimal value of $J(x,y,\vartheta)$ for optimization problem \eqref{sec:Mm.7}
is given by
$$
\max_\zs{\vartheta\in\cV}\,J(x,y,\vartheta)\,=\,J(x,y,\vartheta^{*})\,
=x^\gamma \,(h(0,y))^\varepsilon\,,
$$
where $h(t,y)$ is the unique solution of the equation \eqref{sec:HJB.4}.
Moreover, for all $0\le t\le T$ an optimal financial strategy 
$\vartheta^{*}=(\pi^{*},c^{*})$ is  of the form
\begin{equation}\label{sec:Mr.13}
\left\{
\begin{array}{rl}
\pi^{*}_\zs{t}=\pi^{*}(t,Y_\zs{t})
&=\,\dfrac{\theta(t,Y_\zs{t})}{1-\gamma}
+
\dfrac{\varepsilon\sqrt{1-\rho^{2}}\beta\sigma_\zs{*}D_\zs{y}h(t,Y_\zs{t})}
{(1-\gamma)\,h(t,Y_\zs{t})}
\,;
 \\[5mm]
c^{*}_\zs{t}=
c^{*}(t,Y_\zs{t})
&=
\left(h(t,Y_\zs{t})\right)^{-q_\zs{*}}\,.
\end{array}
\right.
\end{equation}
The optimal wealth process $(X^{*}_t)_{0\le t\le T}$ satisfies
the following stochastic equation
\begin{equation}\label{sec:Mr.14}
\d X^{*}_\zs{t}\,=a^{*}(t,Y_\zs{t}) 
X^{*}_\zs{t}\d t+X^{*}_\zs{t}(b^{*}(t,Y_\zs{t}))'\d W_\zs{t}\,, \quad
\  X^{*}_\zs{0}\,=\,x\,,
\end{equation}
where 
\begin{align*}
a^{*}(t,y)&=
\frac{|\theta(t,y)|^2}{1-\gamma}
+
\frac{\varepsilon\sqrt{1-\rho^{2}}\beta}{(1-\gamma)h(t,y)}
(\sigma_\zs{*}D_\zs{y}h(t,y))'\theta(t,y)
\\[3mm]
&+ r(t,y)-\left( h(t,y)\right)^{-q_\zs{*}}\,; \\[3mm]
b^{*}(t,y)&=
\frac{ \theta(t,y)}{1-\gamma}
+
\frac{\varepsilon\sqrt{1-\rho^{2}}\beta}{(1-\gamma)h(t,y)}\,
\sigma_\zs{*}D_\zs{y}h(t,y)
\,.
\end{align*}
\end{theorem}
\noindent
The proof of this theorem is given in Section~\ref{sec:Pr}.

\bigskip

\begin{remark}\label{Re.sec:Mr.2}
Note that the optimal strategy  \eqref{sec:Mr.13} coincides with the well-known Merton
strategy in the case $\rho=1$, i.e. when the process $Y$ is
independent of the brownian motion $W$ generating 
the financial market \eqref{sec:Mm.1}. Note also that, in the case $0\le \rho<1$, the optimal
investment strategy in \eqref{sec:Mr.13} 
depends on the external factor $Y$ through the derivative
of the function $h$. The first term in this expression is the well-known Merton strategy and
the second term is the impact of the external factor. 
\end{remark}
To calculate the optimal strategy in \eqref{sec:Mr.13} we use the sequence
$(h_\zs{n})_\zs{n\ge 1}$, i.e. we set
$$
\pi^{*}_\zs{n}(t,y)=\frac{\theta(t,y)}{1-\gamma}
+
\frac{\varepsilon\sqrt{1-\rho^{2}}\beta}{(1-\gamma)h_\zs{n}(t,y)}
\sigma_\zs{*}D_\zs{y}h_\zs{n}(t,y)
$$
and
$$
c^{*}_\zs{n}(t,y)=\left(h_\zs{n}(t,y)\right)^{-q_\zs{*}}\,.
$$
\noindent Theorem~\ref{Th.sec:Mr.1}--Theorem~\ref{Th.sec:Mr.2} imply the following result

\begin{theorem}\label{Th.sec:Mr.3}
Assume, that the conditions 
$\A_\zs{1}$)--$\A_\zs{3}$) hold. Then for any $n\ge 1$
$$
\sup_\zs{(t,y)\in \cK}\,
\left(|\pi^{*}(t,y)-\pi^{*}_\zs{n}(t,y)|
+|c^{*}(t,y)-c^{*}_\zs{n}(t,y)|
\right)\le \B^{*}_\zs{1}\,\U^{*}_\zs{n}\,,
$$
where $\B^{*}_\zs{1}=
\gamma_\zs{1}\beta\varepsilon\sqrt{1-\rho^{2}}
|\sigma_\zs{*}|
\r^{*}
+
q_\zs{*}$.

\end{theorem}

\begin{remark}\label{Re.sec:Mr.3}
Note that in this paper we use only the power utility function
$x^{\gamma}$ with $0<\gamma<1$. It seems that the question
about the upper bound  \eqref{sec:Mr.10}
is open for general utility functions of the CRRA type.
The method to obtain the
 bound  \eqref{sec:Mr.10}  is based on the explicite form of the utility function.
The heart  of this method is to show that the operator
\eqref{sec:Mr.5} is contracted in a suitable functional space.
To this end one needs to make use of the power uitility.
\end{remark}

\medskip
\medskip

\subsection{Formula for $H^{*}$}\label{sec:Fr}
 To write the upper bound for the partial derivatives of the 
function $\cH_\zs{f}(t,s,y)$ (see Lemma~\ref{Le.sec:A.1}) we define
for any $q>0$ 
\begin{equation}\label{sec:Fr.1}
\phi_\zs{q}=\sqrt{2}\,e^{T\wt{\phi}_\zs{q}}
\quad\mbox{and}\quad
\iota_\zs{q}=q\alpha_\zs{*}\,(\beta^{-2}+T)\,,
\end{equation}
where 
$
\wt{\phi}_\zs{q}=
q\left(Q_\zs{*}
+\beta^{2}\alpha_\zs{*}/2
\right)
+(2m^{2}+1)q^{2}\beta^{-2}\alpha^{2}_\zs{*}$
.
Using these parameters we set
\begin{equation}\label{sec:Fr.2}
H^{*}=
2 \,
\max\left(
2\phi_\zs{1}
(
\iota_\zs{1}
+\alpha_\zs{*}
)
+
\Psi_\zs{*}\,(\phi_\zs{2})^{1/2}
\,,\,
\frac{1}{\beta\sqrt{2\pi}}
\right)
\,
e^{\frac{\wt{\iota}}{2}}
\,,
 \end{equation}
where 
$\Psi_\zs{*}=
\sqrt{
T\alpha_\zs{*}+
\left(
T
D_\zs{*}+T\alpha_\zs{*}
\left(
\alpha_\zs{*}+1+\beta^{2}/2
\right)
+\beta^{-2}
\alpha_\zs{*}
\right)^{2}}$ and\\
$\wt{\iota}=
\beta^{2}T\,
\left(\max\left(\iota_\zs{1}
\,,\,
\iota_\zs{2}
\right)
+\alpha_\zs{*}
\right)^{2}$.

\medskip

\medskip
\section{Two-asset stochastic volatility model}\label{sec:Sv}

In this section we consider an important example of the model \eqref{sec:Mm.1}
used in financial markets with the stochastic volatilities 
(see, e.g., \cite{Pham2002}).
This model is defined as follows:
$$
\d S_\zs{0}(t)=r\,S_\zs{0}(t)\d t
$$
and 
\begin{align*}
\d S_\zs{1}(t)&=\mu_\zs{1}\,S_\zs{1}(t)\d t
+S_\zs{1}(t)\sigma_\zs{1}(t,Y^{1}_\zs{t})
\left(\d W^{1}_\zs{t}
+
\b_\zs{1}\d W^{2}_\zs{t}\right)
\,;\\[2mm]
\d S_\zs{2}(t)&=\mu_\zs{2}S_\zs{2}(t)\d t+
S_\zs{2}(t)\sigma_\zs{2}(t,Y^{2}_\zs{t})
\left(\d W^{1}_\zs{t}
+
\b_\zs{2}\d W^{2}_\zs{t}\right)\,,
\end{align*}
where $\b_\zs{1}\neq \b_\zs{2}$, 
the process $Y_\zs{t}=(Y^{1}_\zs{t},Y^{2}_\zs{t})'$ is gouverned by the following
stochastic differential equations
\begin{align*}
\d Y^{1}_\zs{t}&=F_\zs{1}(t,Y^{1}_{t})\,\d t+\beta\d \U^{1}_{t}\,,\\[2mm]
\d Y^{2}_\zs{t}&=F_\zs{2}(t,Y^{2}_{t})\,\d t+\beta\d \U^{2}_{t}\,.
\end{align*}
Here $\U^{i}_{t}=\rho V^{i}_\zs{t}+\sqrt{1-\rho^{2}}W^{i}_\zs{t}$
for some $0\le \rho\le 1$, i.e. the parameter $\sigma_\zs{*}=1$ in
\eqref{sec:Mm.2-1}. Assume, 
that the $[0,T]\times\bbr\to\bbr$ functions $F_\zs{i}$ are bounded and have 
 bounded derivatives, i.e.
$$
F_\zs{max}=\max_\zs{1\le i\le 2}\,
\sup_\zs{(t,y)}\,\max\,
\left(
|F_\zs{i}(t,y)|\,,\,
\left|
\frac{\partial  F_\zs{i}(t,y)}{\partial t}
\right|\,,\,
\left|
\frac{\partial  F_\zs{i}(t,y)}{\partial y}
\right|
\right)
<\infty\,.
$$
We assume also, that the 
$[0,T]\times\bbr\to\bbr$ functions 
$\sigma_\zs{i}$ are such, that
$$
\sigma_\zs{max}=\max_\zs{1\le i\le 2}\,
\sup_\zs{(t,y)}\,\max\,
\left(
\left|
\frac{\partial  \sigma_\zs{i}(t,y)}{\partial t}
\right|
\,,\,
\left|
\frac{\partial  \sigma_\zs{i}(t,y)}{\partial y}
\right|
\right)
<\infty
$$
and 
$$
\sigma_\zs{min}=
\inf_\zs{(t,y)}\,
\min\left(
|\sigma_\zs{1}(t,y)|\,,\,|\sigma_\zs{2}(t,y)|
\right)
>0\,.
$$ 
Obviously, that in this case the matrix 
$\sigma(t,y_\zs{1},y_\zs{2})$ is non-degenerated and
$$
\sigma^{-1}(t,y_\zs{1},y_\zs{2})=
\left(
\begin{array}{rl}
\dfrac{\b^{*}_\zs{1}}{\sigma_\zs{1}(t,y_\zs{1})}\,&;\, 
-\dfrac{\b^{*}_\zs{1}}{\sigma_\zs{2}(t,y_\zs{2})}\\[5mm]
-\dfrac{\b^{*}_\zs{2}}{\sigma_\zs{1}(t,y_\zs{1})}\,&;\ \ 
\dfrac{\b^{*}_\zs{2}}{\sigma_\zs{2}(t,y_\zs{2})}
\end{array}
\right)\,,
$$
where $\b^{*}_\zs{1}=\b_\zs{2}(\b_\zs{2}-\b_\zs{1})^{-1}$ and 
$\b^{*}_\zs{2}=(\b_\zs{2}-\b_\zs{1})^{-1}$.
Therefore, the components of the function 
\eqref{sec:Mm.3} are
$$
[\theta(t,y_\zs{1},y_\zs{2})]_\zs{i}=
\frac{\wt{\b}_\zs{1,i}}{\sigma_\zs{1}(t,y_\zs{1})}
 +\frac{\wt{\b}_\zs{2,i}}{\sigma_\zs{2}(t,y_\zs{2})}\,,
\quad 1\le i\le 2\,,
$$
where     
$\wt{\b}_\zs{1,1}=\b^{*}_\zs{1}(\mu_\zs{1}-r)$,
$\wt{\b}_\zs{2,1}=-\b^{*}_\zs{1}(\mu_\zs{2}-r)$,
$\wt{\b}_\zs{1,2}=-\b^{*}_\zs{2}(\mu_\zs{1}-r)$
and 
$\wt{\b}_\zs{2,2}=\b^{*}_\zs{2}(\mu_\zs{2}-r)$.                                  
From this we obtain, that in this case
 function \eqref{sec:Mm.8} has the form 
\eqref{sec:Mm.9}, i.e. 
$$
[\alpha(t,y_\zs{1},y_\zs{2})]_\zs{i}=F_\zs{i}(t,y_\zs{i})
+
\beta_\zs{*}
\left(
\frac{\wt{\b}_\zs{1,i}}{\sigma_\zs{1}(t,y_\zs{1})}
 +\frac{\wt{\b}_\zs{2,i}}{\sigma_\zs{2}(t,y_\zs{2})}
\right)
\,,\quad 1\le i\le 2
\,.
$$
It is easy to see, that for this model
 the conditions $\A_\zs{1}$)--$\A_\zs{3}$)
hold with
$$
\alpha_\zs{*}\le
F_\zs{max}
+
\wt{\b}_\zs{*}
\,
\frac{1+\sigma_\zs{max}}{\sigma_\zs{min}}
\quad\mbox{and}\quad
\wt{\b}_\zs{*}=\beta_\zs{*}\,\max_\zs{1\le i,j\le 2} (|\wt{\b}_\zs{1,i}|)\,.
$$

\medskip
\medskip

\section{Properties of the mapping $\cL$}\label{sec:PrL}

\begin{proposition}\label{Pr.sec:PrL.2} Assume, that the conditions
$\A_\zs{1}$)--$\A_\zs{3}$) hold. Then $\cL_\zs{f}\in\cX$
 for any $f\in \cX $.
\end{proposition}

\proof
Obviously, that for any $f\in\cX$ the mapping $\cL_\zs{f} \ge 1$. Moreover,
 setting
\begin{equation}\label{sec:PrL.1}
\wt{f}_\zs{s}=f(s,\eta^{t,y}_\zs{s})
\,,
\end{equation}
we represent $\cL_\zs{f}(t,y)$ as
\begin{equation}\label{sec:PrL.2}
\cL_\zs{f}(t,y) =
\E\,\cG(t,T,y)
+\frac{1}{q_\zs{*}}
\, \int_{t}^{T}
\E \left(
\wt{f}_\zs{s}\right) ^{1-q_\zs{*}}\,
\cG(t,s,y) ds\,.
\end{equation}
Therefore, taking into account that $\wt{f}_\zs{s}\ge 1$
and $q_\zs{*}\ge 1$ we get
\begin{equation}\label{sec:PrL.3}
\cL_\zs{f}(t,y)
\,
\le
\,
e^{Q_\zs{*}(T-t)}+\int_{t}^{T}\dfrac{1}{q_\zs{*}}\,
e^{Q_\zs{*}(s-t)}\d s
\,\le \,\r_\zs{0}\,,
\end{equation}
where the upper bound $\r_\zs{0}$ is defined in \eqref{sec:Fr.4}.
\noindent Moreover, Lemmas~\ref{Le.sec:A.3}--\ref{Le.sec:A.4} yield
$$
\frac{\partial }{\partial y_\zs{i}}
\cL_\zs{f}(t,y)
= \E
\frac{\partial }{\partial y_\zs{i}}
\cG(t,T,y)
+\frac{1}{q_\zs{*}}\,
\int_{t}^{T}\frac{\partial }{\partial y_\zs{i}}
\cH_\zs{f}(t,s,y)\,\d s \,.
$$
\noindent 
Therefore,  
applying here \eqref{sec:PrL.3}, \eqref{sec:A.18} and  Lemmas~\ref{Le.sec:A.1}
we get
$$
\cL_\zs{f}(t,y)
+
\left|
\D_\zs{y}\cL_\zs{f}(t,y)
\right|
\le \r^{*}\,.
$$
Now we have to  show that the function $\cL_\zs{f}(t,y)$
is continuously differentiable with respect to $t$ for any $f\in\cX$. Indeed,
to this end we consider for any $f$ from $\cX$
the  equation \eqref{sec:HJB.4}, i.e.
\begin{equation}\label{sec:PrL.4}
\left\{
\begin{array}{rl}
u_\zs{t}(t,y)&+
Q(t,y) u(t,y)
+
(\alpha(t,y))'\D_\zs{y}\,u(t,y)\\[5mm]
&
+\dfrac{\beta^{2}}{2}\tr \D_\zs{y,y} u(t,y)
+\dfrac{1}{q_\zs{*}}
\left(\dfrac{1}{f(t,y)}\right)^{q_\zs{*}-1}
=0\,;
\\[5mm]
u(T,y)&=1\,.
\end{array}
\right.
\end{equation}
Setting here $\wt{u}(t,y)=u(T-t,y)$ we obtain a uniformly parabolic equation
for $\wt{u}$ with initial condition $\wt{u}(0,y)=1$. Moreover, due to the conidtions 
$\A_\zs{1}$) and   $\A_\zs{3}$) the functions $Q$ and $\alpha$ have bounded 
partial derivatives with respect to $y$. Therefore, for any $f$ from $\cX$
Theorem 5.1 from \cite{LadyzenskajaSolonnikovUralceva1967} (p. 320)
with  $0<l<1$  provides the existence of the unique solution of \eqref{sec:PrL.4}
belonging to $\C^{1,2}(\cK)$. 
Applying the It\^o formula to the process
$$
\left(u(s,\eta^{t,y}_\zs{s})\,
e^{\int^{s}_\zs{t} Q(v,\eta^{t,y}_\zs{v})\d v}\right)_\zs{t\le s\le T} 
$$
 and taking into account the equation \eqref{sec:PrL.4}
we get
\begin{equation}\label{sec:PrL.5}
u(t,y)=\cL_\zs{f}(t,y)\,.
\end{equation}

\noindent 
Therefore, the function
$\cL_\zs{f}(t,y)\in \C^{1,2}(\cK)$, i.e. $\cL_\zs{f}\in\cX$
for any $f\in\cX$. Hence Proposition~\ref{Pr.sec:PrL.2}.
\endproof

\bigskip

\begin{proposition}\label{Pr.sec:Prl.3}
Under the conditions $\A_\zs{1})$--$\A_\zs{3})$ 
the mapping $\cL$ is a contraction in the metric space
$(\cX , \varrho_\zs{*})$, i.e.
for any $f$, $g$ from $\cX$
\begin{equation}\label{sec:PrL.6}
\varrho_\zs{*}(\cL_\zs{f},\cL_\zs{g})
\le \lambda \varrho_\zs{*}(f,g)\,,
\end{equation}
where the parameter $0<\lambda<1$ is given in \eqref{sec:Mr.9}.
\end{proposition}
\proof
First note that,
for any $f$ and $g$ from $\cX$ and for any $y\in\bbr^{m}$
\begin{align*}
|\cL_\zs{f}(t,y)-\cL_\zs{g}(t,y)|
&\le
\frac{1}{q_\zs{*}}\,
 \E\,
\int_{t}^{T}
\cG(t,s,y)
\left\vert
\left(\wt{f}_\zs{s}\right) ^{1-q_\zs{*}}
-
\left(\wt{g}_\zs{s}\right) ^{1-q_\zs{*}}
\right\vert
\d s\\[2mm]
&\le
\,
\int^{T}_\zs{t}\,
\E\,\cG(t,s,y)
\,\left\vert
\wt{f}_\zs{s}
-
\wt{g}_\zs{s}
\right\vert
\d s  \,.
\end{align*}
We recall that  $\wt{f}_\zs{s}=f(s,\eta^{t,y}_\zs{s})$
and $\wt{g}_\zs{s}=g(s,\eta^{t,y}_\zs{s})$. Taking into account here that
$\cG(t,s,y)\le e^{Q_\zs{*}(s-t)}$ we obtain
$$
|
\cL_\zs{f}(t,y)-\cL_\zs{g}(t,y)
|
\le
\,
\int^{T}_\zs{t}\,e^{Q_\zs{*}(s-t)}
\,
\E
|
\wt{f}_\zs{s}-\wt{g}_\zs{s}
|
\d s
\,.
$$
Taking into account in the last inequality, that
\begin{equation}\label{sec:PrL.7}
|
\wt{f}_\zs{s}-\wt{g}_\zs{s}
|
\le e^{\varkappa (T-s)}\,\varrho_\zs{*}(f,g)
\quad\mbox{a.s.}
\,,
\end{equation}
we get
\begin{equation}\label{sec:PrL.8}
\left\vert e^{-\varkappa(T-t)}
\left( \cL_\zs{f}(t,y)-\cL_\zs{g}(t,y)\right) \right\vert
\le
\frac{1}{\varkappa-Q_\zs{*}}\,\varrho_\zs{*}(f,g)
\,.
\end{equation}
Moreover, by virtue of Lemma~\ref{Le.sec:A.3} we obtain, that for any $1\le i\le m$
\begin{align}\nonumber
\frac{\partial }{\partial y_\zs{i}}
\cL_\zs{f}(t,y)
&=
\E\,
\frac{\partial }{\partial y_\zs{i}}
\cG(t,T,y)\\[2mm]\nonumber
&+\frac{1-q_\zs{*}}{q_\zs{*}}\,
\int_{t}^{T}
\,\E\,
(\wt{f}_\zs{s})^{-q_\zs{*}}\,
\left(\wt{f}_\zs{0}(s)\right)'\,
\upsilon_\zs{i}(s)
\cG(t,s,y)
\d s\\[2mm]\label{sec:PrL.9}
&+\frac{1}{q_\zs{*}}\,
\int_{t}^{T}
\,\E\,
(\wt{f}_\zs{s})^{1-q_\zs{*}}\,
\frac{\partial }{\partial y_\zs{i}}
\cG(t,s,y)
\d s
 \,,
\end{align}
where $\wt{f}_\zs{0}(s)=f_\zs{0}(s,\eta^{t,y}_\zs{s})$, 
$f_\zs{0}(s,z)=\D_\zs{z}\,f(s,z)$ 
and $\upsilon_\zs{i}(s)=\partial \, \eta_\zs{s}^{t,y} / \partial \, y_\zs{i}$.
Therefore, for any $f$ and $g$ from $\cX$
\begin{align*}
\frac{\partial \cL_\zs{f}(t,y)}{\partial y_\zs{i}}
-
\frac{\partial \cL_\zs{g}(t,y)}{\partial y_\zs{i}}
&=
\frac{1-q_\zs{*}}{q_\zs{*}}\,
\int_{t}^{T}
\,\E\,
\left(\varpi_\zs{1}(s)\right)'
\upsilon_\zs{i}(s)
\cG(t,s,y)
\d s\\[2mm]
&+\frac{1}{q_\zs{*}}\,
\int_{t}^{T}
\,\E\,
\varpi_\zs{2}(s)
\,
\frac{\partial }{\partial y_\zs{i}}
\cG(t,s,y)
\d s
 \,,
\end{align*}
where
$$
\varpi_\zs{1}(s)=
\frac{\wt{f}_\zs{0}(s)}{(\wt{f}_\zs{s})^{q_\zs{*}}}
-
\frac{\wt{g}_\zs{0}(s)}{(\wt{g}_\zs{s})^{q_\zs{*}}}
\quad\mbox{and}\quad
\varpi_\zs{2}(s)=
(\wt{f}_\zs{s})^{1-q_\zs{*}}
-
(\wt{g}_\zs{s})^{1-q_\zs{*}}\,.
$$
Note, that similarly to \eqref{sec:PrL.7}
$$
|
\wt{f}_\zs{0}(s)-\wt{g}_\zs{0}(s)
|
\le e^{\varkappa (T-s)}\,\varrho_\zs{*}(f,g)
\quad\mbox{a.s.}
\,.
$$
Using this, one can check directly that
$$
|\varpi_\zs{1}(s)|\le (1+\r^{*}q_\zs{*})e^{\varkappa (T-s)}\,
\varrho_\zs{*}(f,g)
$$
and
$$
|\varpi_\zs{2}(s)|\le (q_\zs{*}-1)e^{\varkappa (T-s)}\,
\varrho_\zs{*}(f,g)\,.
$$
Therefore, these upper bounds and Lemma~\ref{Le.sec:A.2} imply
$$
\left|
\frac{\partial \cL_\zs{f}(t,y)}{\partial y_\zs{i}}
-
\frac{\partial \cL_\zs{g}(t,y)}{\partial y_\zs{i}}
\right|
\le
e^{\varkappa (T-t)}
\frac{\L^{*}
}{\varkappa-Q_\zs{*}}\,\varrho_\zs{*}(f,g)\,,
$$
where $\L^{*}$
is given in \eqref{sec:Mr.2}.
Therefore,
$$
\sup_\zs{(t,y)\in \cK}
e^{-\varkappa(T-t)}
\left|
\D_\zs{y} \cL_\zs{f}(t,y)
-
\D_\zs{y} \cL_\zs{g}(t,y)
\right|
\le
\frac{m\L^{*}}{\varkappa-Q_\zs{*}}\,\varrho_\zs{*}(f,g)\,.
$$
Taking into account the definition of $\varkappa$ in 
\eqref{sec:Mr.2}, 
we obtain the inequality \eqref{sec:PrL.6}. 
Hence Proposition~\ref{Pr.sec:Prl.3}. \endproof

\begin{proposition}\label{Pr.sec:Prl.4}
Under the conditions $\A_\zs{1})$--$\A_\zs{3})$ 
the equation $\cL_\zs{h}=h$
has a uniqe solution in $\cX$.
\end{proposition}
\proof
Indeed, using the contraction of the operator $\cL$ in 
$\cX$ and the definition of the sequence $(h_\zs{n})_\zs{n\ge 1}$
in \eqref{sec:Mr.8} we get, that for any $n\ge 1$
\begin{equation}\label{sec:PrL.10}
\varrho_\zs{*}(h_\zs{n},h_\zs{n-1})\le \lambda^{n-1}\,
\varrho_\zs{*}(h_\zs{1},h_\zs{0})\,,
\end{equation}
i.e. the sequnce \eqref{sec:Mr.8} is fundamental in $(\cX,\varrho_\zs{*})$.
Therefore, due to Proposition~\ref{Pr.sec:PrL.1} 
this sequence has a limit in $\cX$, i.e.
there exits a function $h$ from $\cX$ for which
$$
\lim_\zs{n\to\infty}\,\varrho_\zs{*}(h,h_\zs{n})=0\,.
$$
Moreover, taking into account that 
$h_\zs{n}=\cL_\zs{h_\zs{n-1}}$ we obtain, that
for any $n\ge 1$
$$
\varrho_\zs{*}(h,\cL_\zs{h})
\le 
\varrho_\zs{*}(h,h_\zs{n})
+
\varrho_\zs{*}(\cL_\zs{h_\zs{n-1}},\cL_\zs{h})
\le \varrho_\zs{*}(h,h_\zs{n})
+\lambda
\varrho_\zs{*}(h,h_\zs{n-1})\,.
$$
The last expression tends to zero as $n\to\infty$. Therefore 
$\varrho_\zs{*}(h,\cL_\zs{h})=0$, i.e. $h=\cL_\zs{h}$. 
Propostion~\ref{Pr.sec:Prl.3} implies immediately that this solution is unique.
\endproof

\begin{proposition}\label{Pr.sec:Prl.5}
Under the conditions $\A_\zs{1})$--$\A_\zs{3})$ 
the equation \eqref{sec:HJB.4} has a unique solution which is the solution
$h$ of the fixed-point equation $\cL_\zs{h}=h$.
\end{proposition}
\proof
Choosing in \eqref{sec:PrL.4} the function $f=h$
and taking into account
the representation \eqref{sec:PrL.5} and the equation \eqref{sec:Mr.7}
we obtain, that the solution of the equation \eqref{sec:PrL.4}
$$
u=\cL_\zs{h}=h\,.
$$
Therefore, the function $h$ satisfies the equation \eqref{sec:HJB.4}.
Moreover, this solution is unique since $h$ is the unique solution
of the equation  \eqref{sec:Mr.7}.
\endproof

\begin{remark}\label{Re.sec:PrL.1}
The representation \eqref{sec:PrL.5}
for the solution of the equation  \eqref{sec:PrL.4}
is called the probabilistic representation or
the Feynman - Kac formula (see, e. g., \cite{KabanovPergamenshchikov2003}, p. 194).
\end{remark}

\section{Verification theorem }\label{sec:Vt}

In this section we state  the verification theorem from
\cite{KluppelbergPergamenchtchikov2009}.
 Consider on the interval $[0,T]$ the stochastic control process given by the
$N$ - dimensional It\^{o} process

\begin{equation} \label{sec:A.1}
\left\{
\begin{array}{l}
\d \varsigma_\zs{t}^{\vartheta }\,=\a(t,\varsigma_\zs{t}^{\vartheta },\vartheta _{t})\,\d t\,+\b(t,\varsigma_\zs{t}^{\vartheta },
\vartheta _{t})\,\d W_{t}\,,\quad t\geq 0\,, \\[2mm]
\varsigma_\zs{0}^{\vartheta}=x\in\bbr^{N}\,,
\end{array}
\right.
\end{equation}
where $(W)_\zs{0\le t\le T}$ is a standard $k$ - dimensional Brownian motion.
We assume that the control
process $\vartheta $ takes values in some set $\Theta$. Moreover, 
we  assume that the coefficients $\a$
and $\b$ satisfy the following conditions

\begin{itemize}
\item  for all $t\in \lbrack 0,T]$ the functions $\a(t,\cdot,\cdot)$
and $\b(t,\cdot,\cdot )$ are continuous on $ \bbr^{N} \times  \Theta$;

\item  for every deterministic vector $\upsilon \in  \Theta$ the
stochastic differential equation
\begin{equation*}
\d \varsigma_\zs{t}^{\vartheta }=\a(t,\varsigma_\zs{t}^{\vartheta },\vartheta )\,%
\d t\,+\b(t,\varsigma_\zs{t}^{\vartheta },\vartheta )\,
\d W_{t}
\end{equation*}
has a unique strong solution.
\end{itemize}

\noindent Now we introduce admissible control processes for the equation \eqref{sec:A.1}.
We set $\mathcal{F}_{t}=\sigma\{W_{u}\,,0\le u\le t\}$ for any $0<t\le T$.

\begin{definition}\label{De.sec:A.1}
 A stochastic control process $\vartheta =(\vartheta _{t})_{0\le t\le T}$
is called \emph{admissible} on $[0,T]$ with respect to equation
\eqref{sec:A.1} if it is $(\mathcal{F}_{t})_{ 0\le t\le T}$ - progressively measurable with values in
$ \Theta$, and equation \eqref{sec:A.1} has a unique strong a.s.
continuous solution $(\varsigma_\zs{t}^{\vartheta})_{ 0\le t\le T}$  such that
\begin{equation}\label{sec:A.2}
\int_{0}^{T}\,\left( |\a(t,\varsigma_\zs{t}^{\vartheta},\vartheta _{t})|+
|\b(t,\varsigma_\zs{t}^{\vartheta },\vartheta _{t})|^{2}\right)
\d t\,<\,\infty \quad \mbox{a.s..}
\end{equation}
\end{definition}
\noindent We denote by $\cV$  the set of all admissible control processes
with respect to the equation \eqref{sec:A.1}.

Moreover, let
$\f\,:[0,T]\,\times {\bbr}^{m}\times \, \Theta\,\rightarrow \,[0,\infty )$ and
$\h\,:\,\bbr^{m}\,\rightarrow \,[0,\infty )$ be continuous
utility functions. We define the cost function by
$$
J(t,x,\vartheta)=\E_{t,x}\,
\left(
\int_{t}^{T}\f(s,\varsigma_\zs{s}^{\vartheta },\vartheta _{s})\,\d s+\h(\varsigma_\zs{T}^{\vartheta})
\right)
\,,\quad 0\leq t\leq T\,,
$$
where $\mathbf{E}_{t,x}$ is the expectation
operator conditional on $\varsigma_\zs{t}^{\vartheta}=x $. Our goal is to solve the optimization problem
\begin{equation} \label{sec:A.3}
J^{*}(t,x)\,:=\,\sup_{\vartheta\in\cV}\,J(t,x,\vartheta )\,. 
\end{equation}
To this end we introduce the Hamilton function, i.e. for any $0\le t\le T$,
$\varsigma,\q\in\bbr^{N}$ and symmetric $N\times N$ matrix $\M$ we set
\begin{equation}\label{sec:A.4}
H(t,\varsigma,\q,\M)\,:=\,\sup_{\vartheta\in \Theta}\,H_{0}(t,\varsigma,\q,\M,\vartheta )\,,
\end{equation}
where
$$
H_\zs{0}(t,\varsigma,\q,\M,\vartheta):=\a^{\prime}(t,\varsigma,\vartheta )\q+\,
\frac{1}{2}\tr\left( \b\b^{\prime}(t,\varsigma,\vartheta)\M\right) +
\f(t,\varsigma,\vartheta )\,.
$$
In order to find the solution to %
\eqref{sec:A.3} we investigate the \emph{Hamilton-Jacobi-Bellman} equation
\begin{equation} \label{sec:A.5}
\left\{
\begin{array}{ll}
z_{t}(t,\varsigma)+
\,H(t,\varsigma,\D_\zs{\varsigma}z(t,\varsigma),\D_\zs{\varsigma,\varsigma}z(t,\varsigma))=0\,,\,
& \ t\in [ 0,T]\,, \\[5mm]
z(T,\varsigma)=\h(\varsigma)\,,\,  & \varsigma\in\bbr^{N}\,.
\end{array}
\right.
\end{equation}

Here $z_{t}$ denotes the partial derivative of $z$ with respect to $t$, 
$D_\zs{\varsigma} z(t,\varsigma)$ the gradient vector with respect to
$\varsigma$ in $\bbr^{N}$ and $D_\zs{\varsigma,\varsigma} z(t,\varsigma)$ denotes
the symmetric hessian matrix, that is the matrix of the second order partial
derivatives with respect to $\varsigma$.

We assume that the following conditions hold:

\noindent$\H_\zs{1})$ {\em The functions $\f$ and $\h$ are non negative.}

\medskip

\noindent$\H_\zs{2})$
{\em There exists
 $[0,T]\times \bbr^{N}\rightarrow (0,\infty )$
function $z(t,\varsigma)$ from

 $\C^{1,2}([0,T]\times \bbr^{N})$
which satisfies the HJB equation \eqref{sec:A.5}.
}

\medskip

\noindent $\H_\zs{3})$ {\em There exists a measurable function
$\vartheta^{*}\,:\,[0,T]\times \bbr^{N} \to  \Theta$
 such that for all $0\le t\le T$ and $\varsigma\in\bbr^{N}$}
$$
H(t,\varsigma,\D_\zs{\varsigma}z(t,\varsigma),\D_\zs{\varsigma,\varsigma}z(t,\varsigma))
=
H_\zs{0}
(t,\varsigma,\D_\zs{\varsigma}z(t,\varsigma),\D_\zs{\varsigma,\varsigma}z(t,\varsigma),
\vartheta^{*}(t,\varsigma))\,.
$$

\medskip

\noindent $\H_\zs{4})${\em There exists a unique strong solution to the It\^{o}
equation
\begin{equation}\label{sec:A.6}
\d \varsigma^{*}_\zs{t}=\a^{*}(t,\varsigma^{*}_\zs{t})\,\d t+
\b^{*}(t,\varsigma^{*}_\zs{t})\,\d W_\zs{t}\,,\quad t\ge 0\,,\quad
 \varsigma^{*}_\zs{0}\,=\,x\,,
\end{equation}
where $\a^{*}(t,\cdot)=\a(t,\cdot,\vartheta^{*}(t,\cdot))$ and
$\b^{*}(t,\cdot)=\b(t,\cdot,\vartheta^{*}(t,\cdot))$.
Moreover, the optimal control process
$\vartheta^{*}_\zs{t}=\vartheta^{*}(t,\varsigma^{*}_\zs{t})$ for $0\le t\le T$ belongs to $\cV$.}

\medskip

\noindent $\H_\zs{5})$ {\em There exists some $\delta>1$ such that
for all $0\le t\le T$ and $\varsigma\in\bbr^{N}$
$$
\sup_{\tau\in \cT_\zs{t}}\E_\zs{t,\varsigma}\,(z(\tau,\varsigma^{*}_\zs{\tau}))^\delta\,
<\,\infty\,
$$
where $\cT_{\zs{t}}$ is the set of all stopping times in $[t,T]$.
}

\begin{theorem}\label{Th.sec:A.1}
Assume that $\cV\ne\emptyset$ and $\H_\zs{1})-\H_\zs{5})$ hold.
Then for all $t\in [0,T]$ and for all $x\in\bbr^{N}$ the solution of the
Hamilton-Jacobi-Bellman equation
\eqref{sec:A.5} coincides with the optimal value of the cost function,
i.e.
$$
z(t,x)=J^{*}(t,x)=J^{*}(t,x,\vartheta^{*})\,,
$$
where the optimal strategy
$\vartheta^{*}$ is defined in  $\H_\zs{3})$ and $\H_\zs{4})$.
\end{theorem}

\bigskip
\section{Proofs}\label{sec:Pr}

\subsection{Proof of Theorem~\ref{Th.sec:Mr.1}}
Proposition~\ref{Pr.sec:Prl.4} implies 
the first part of this theorem. Moreover, from \eqref{sec:PrL.10}
it is easy to see, that
for each $n\ge 1$
$$
\varrho_\zs{*}(h,h_\zs{n})\le \frac{\lambda^{n}}{1-\lambda}\,
\varrho_\zs{*}(h_\zs{1},h_\zs{0})\,.
$$
Thanks to Proposition~\ref{Pr.sec:PrL.2} all the functions
$h_\zs{n}$ belong to $\cX$, i.e.
by the definition of the space $\cX$
$$
\varrho_\zs{*}(h_\zs{1},h_\zs{0})\le
\sup_\zs{(t,y)\in\cK}\,
\left(
|h_\zs{1}(t,y)-1|
+|\D_\zs{y}h_\zs{1}(t,y)|
\right)
\le 1+\r^{*}\,.
$$
Taking into account that
$$
\sup_\zs{(t,y)\in \cK}
\left(
|\Delta_\zs{n}(t,y)|
+
\left|
\D_\zs{y} \Delta_\zs{n}(t,y)
\right|
\right)
\le e^{\varkappa T}
\varrho_\zs{*}(h,h_\zs{n})\,,
$$
we obtain the inequality \eqref{sec:Mr.10}. Hence
Theorem~\ref{Th.sec:Mr.1}. \endproof

\subsection{Proof of Theorem~\ref{Th.sec:Mr.2}}

We apply the Verification Theorem \ref{Th.sec:A.1} to
Problem~\ref{sec:Mm.7} for the stochastic control differential equation \eqref{sec:Mm.5}.
For fixed $\vartheta=(\pi,c)$, where $\pi\in\bbr^d$ and
$c\in [0,\infty)$, the coefficients in  model \eqref{sec:A.1} are defined as
\begin{align*}
\a(t,\varsigma,\vartheta)&=
\left(x\,(r(t,y)+\pi'\theta(t,y)-c),F(t,y)\right)' \\[2mm]
\b(t,\varsigma,\vartheta)&=\left(
                      \begin{array}{cc}
                        \,x\,\pi'\,;& \0'_\zs{m} \\[2mm]
                        \beta\sqrt{1-\rho^{2}}\sigma_\zs{*}\,;
 & \, \beta\rho\,I_\zs{m} \\
                      \end{array}
                    \right)\,,
                    \end{align*}
where
$\varsigma=(x,y)'\in\bbr^{N}$, $N=m+1$, $k=d+m$,
 $\0_\zs{m}=(0,\ldots,0)'\in\bbr^{m}$, $I_\zs{m}$
is the identity matrix of the order $m$.
Note that
$$
\b\b'(t,\varsigma,\vartheta)=\left(
                      \begin{array}{cc}
                        \,x^{2}\,|\pi|^{2}\,;&
 x\beta\sqrt{1-\rho^{2}}\pi'\sigma'_\zs{*}
 \\[2mm]
                        x\beta\sqrt{1-\rho^{2}}\sigma_\zs{*}\,\pi\,; &
 \,\beta^{2}I_\zs{m} \\
                      \end{array}
                    \right)\,.
$$
Therefore, according to the definition of $H_\zs{0}$ in \eqref{sec:A.4},
 for any $\q$ and $\M$ of the form
\eqref{sec:HJB.0}
\begin{align*}
H_\zs{0}(t,\varsigma,&\q,\M,\vartheta)=
x\, r(t,y)\q_\zs{1}+(F(t,y))^{'}\wt{\q}
+(x^{\gamma}c^{\gamma}-xc \q_\zs{1})
\\[2mm]
&+\frac{1}{2} x^{2}\mu|\pi|^{2}
+
x\pi'\left(
\theta(t,y)\q_\zs{1}
+\beta
\sqrt{1-\rho^{2}}\sigma'_\zs{*}\wt{\mu}
\right)
+\frac{\beta^{2}}{2}\tr \M_\zs{0}\,.
\end{align*}
\noindent 
To check the conditions  $\H_\zs{2})-\H_\zs{4})$ we need to calculate
the Hamilton function \eqref{sec:A.4} for Problem~\ref{sec:Mm.7} which is defined as
$$
H(t,\varsigma,\q,\M)=\sup_\zs{\vartheta\in\bbr^d\times [0,\infty)}\,
H_\zs{0}(t,\varsigma,\q,\M,\vartheta)
=
H_\zs{0}(t,\varsigma,\q,\M,\vartheta_\zs{0})
\,.
$$
One can check directly, that for $\mu<0$ and $\q_\zs{1}>0$ we obtain the form \eqref{sec:HJB.1},
where the optimal function
\\
$\vartheta_\zs{0}=\vartheta_\zs{0}(t,\varsigma,\q,\M)=
(\pi^{*}(t,\varsigma,\q,\M),c^{*}(t,\varsigma,\q,\M))$
and
\begin{equation}\label{sec:Pr.1}
\left\{
\begin{array}{rl}
\pi^{*}=
\pi^{*}(t,\varsigma,\q,\M)
&=\,\dfrac{\theta(t,y)\q_\zs{1}
+\beta
\sqrt{1-\rho^{2}}\sigma'_\zs{*}\wt{\mu}}{x|\mu|}\,;
 \\[4mm]
c^{*}=
c^{*}(t,\varsigma,\q,\M)&=
\dfrac{1}{x}\,\left(\dfrac{\gamma}{\q_\zs{1}}\right)^{\gamma_\zs{1}}\,.
\end{array}
\right.
\end{equation}
Proposition~\ref{Pr.sec:Prl.5} implies that the HJB equation \eqref{sec:HJB.2}
has a solution from $\C^{1,2}(\cK)$ defined in  \eqref{sec:HJB.3}. 
It should be noted that for this
function 
$\q_\zs{1}=
z_\zs{x}(t,x,y)=\gamma x^{\gamma-1}h(t,y)>0$
and \\
$
\mu=
z_\zs{xx}(t,x,y)=\gamma(\gamma-1) x^{\gamma-1}h(t,y)<0$.
Therefore, in this case the condition $\H_\zs{3}$) holds
with $\vartheta^{*}(t,\varsigma)=(\pi^{*}(t,\varsigma), c^{*}(t,\varsigma))$
and
\begin{equation}\label{sec:Pr.3}
\left\{
\begin{array}{rl}
\pi^{*}(t,\varsigma)&= \dfrac{\theta(t,y)}{1-\gamma}
+\varepsilon\sqrt{1-\rho^{2}}\beta
\dfrac{\sigma_\zs{*}\D_\zs{y}h(t,y)}{(1-\gamma)h(t,y)} \,;
 \\[5mm]
c^{*}(t,\varsigma)&= \left(h(t,y)\right)^{-q_\zs{*}}\,,
\end{array}
\right.
\end{equation}
where the function $h$ is solution of the equation \eqref{sec:HJB.4}.

Now we check the conditions $\H_\zs{4})$ and $\H_\zs{5})$. 
Note that the equation \eqref{sec:A.6} is identical
to the equations
\eqref{sec:Mm.2} and \eqref{sec:Mm.5}. Due to the condition $\A_\zs{1})$ the coefficients
$a^{*}(t,y)$ and $b^{*}(t,y)$ are bounded. Therefore, the equation  \eqref{sec:Mm.2} has
a unique strong solution which for $0\le t\le s\le T$ can be represented as
$$
X^{*}_\zs{s}=X^{*}_\zs{t}\,e^{\int^{s}_\zs{t} a^{*}(v,Y_\zs{v})\d v}\,\cE_\zs{t,s}\,,
$$
where
$$
\cE_\zs{t,s}=\exp\left\{\int^{s}_\zs{t} (b^{*}(v,Y_\zs{v}))'\d W_\zs{v}-
\frac{1}{2}\int^{s}_\zs{t} |b^{*}(v,Y_\zs{v})|^{2}\d v\right\}\,.
$$
It is clear, that for the bounded function $b^{*}(v,y)$ the process 
$(\cE_\zs{t,s})_\zs{t\le s\le T}$ is a quadratic intergrable martingale and 
by the Doob inequality
$$
\E\,\sup_\zs{t\le s\le T}\,\cE^{2}_\zs{t,s}\,
\le\,4\,\E\,\,\cE^{2}_\zs{t,T}<\infty\,.
$$
This implies directly, that for any $0\le t\le T$, $x>0$ and $y\in\bbr^{m}$
$$
\E
\left(
\sup_\zs{t\le s\le T}\,X^{*}_\zs{s}
|
X^{*}_\zs{t}=x\,,\,
Y_\zs{t}=y
\right)
<\infty\,.
$$
Now we recall that
$$
z(t,\varsigma)= x^\gamma (h(t,y))^\varepsilon
$$
where $h(t,y)$ is some positive function bounded by $\r^{*}$. Thus
\begin{align*}
\E_\zs{t,\varsigma}\,z^{m}(\tau,X^{*}_\zs{\tau},Y_\zs{\tau})
\le (\r^{*})^{m\varepsilon}\,\E_\zs{t,\varsigma}\,
 (X^{*}_\zs{\tau})^{\gamma m}\,.
\end{align*}
Therefore, taking $m = 1/\gamma \, >1 $ one gets
$$
\E_\zs{t,\varsigma}\,z^m(\tau,X^{*}_\zs{\tau},Y_\zs{\tau})
\le (\r^{*})^{\varepsilon/\gamma}\,
\E_\zs{t,\varsigma}\sup_\zs{t\le s\le T} X^{*}_\zs{s}\,<\,\infty
$$
which implies the conditions $\H_\zs{4})$ and  $\H_\zs{5})$.
 Therefore, thanks to Theorem~\ref{Th.sec:A.1} we get Theorem~\ref{Th.sec:Mr.2}.
\endproof

\bigskip

\section{Numerical example}\label{sec:Nu}

In this section through Scilab  we calculate
 the function $h(t,y)$  using the sequence \eqref{sec:Mr.8}
with $n=14$ iterations.
 The curve is obtained
in the following stochastic volatility market settings: the market consists on
one riskless asset (the bond) and a risky one
(that means $d=1$). Moreover, we set  $m=1$, $T=1$,
$$
r(t,y)=0.01 (1+0.5 \sin(y t))\,,\quad
\mu(t,y)=0.02 (1+0.5 \sin(y t))
$$
and
$\sigma(t,y)=0.5+\sin^2(y t)$.
The parameters of the economic factor are
$F(t,y)=0.1 \sin(y t)$, $\beta=1$ and $\rho=0.5$.
The utility function parameter is $\gamma=0,75$.


\begin{center}

    \includegraphics[height=60mm,bb=25 169 579 609]{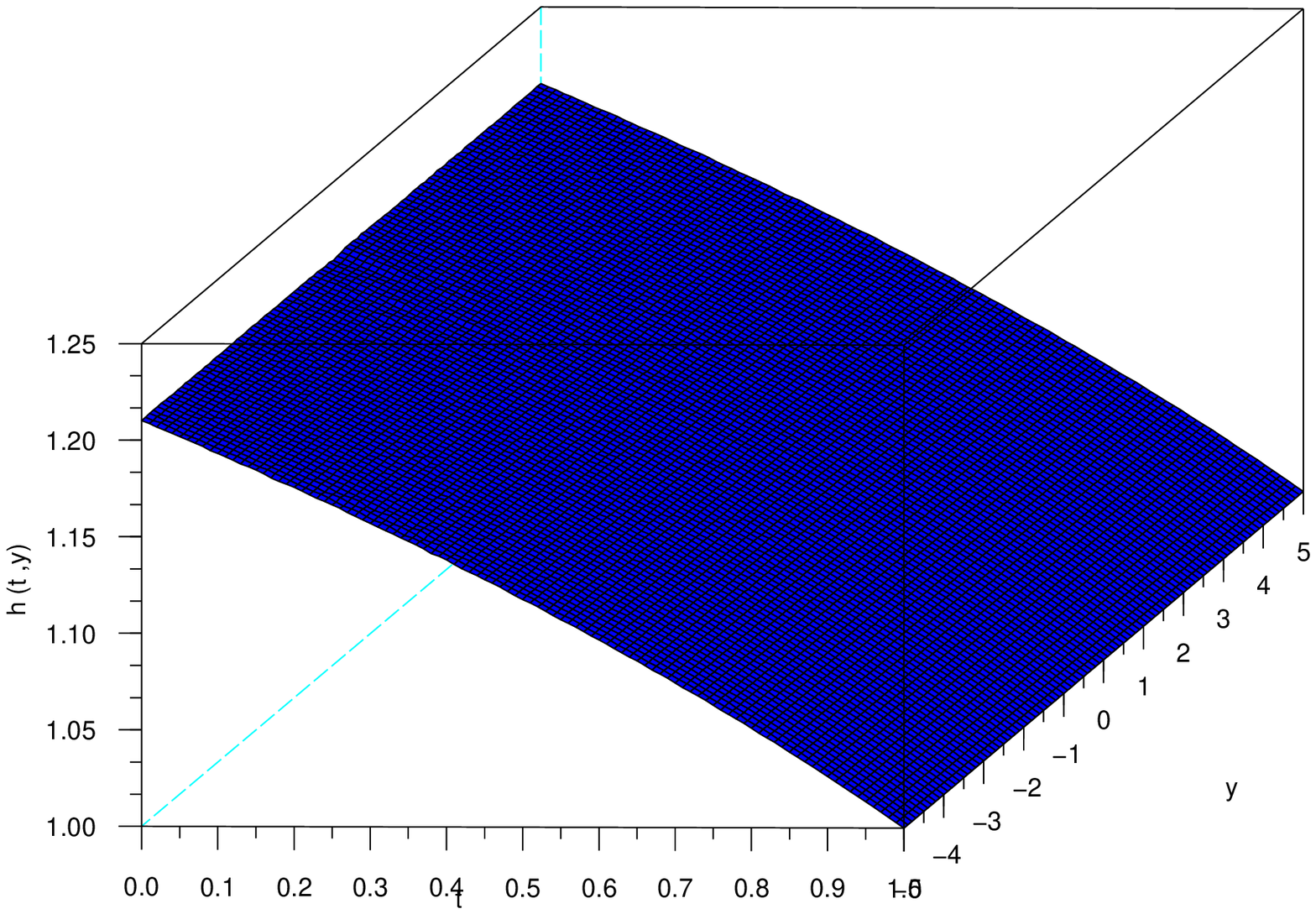}

\textbf{Figure 1: The function h(t,y)}

\end{center}

\vspace{10mm}

The second figure illustrates 
the super geometrical convergence rate for the functions
 $(h_\zs{n})_\zs{n\ge 0}$.
 The curve represents the accuracy $\delta_\zs{n}$ calculated at each step by
 $$
 \delta_\zs{n}=\sup_{(t,y)} |h_\zs{n}(t,y)-h_\zs{n-1}(t,y)|\,.
 $$


\begin{center}

   \includegraphics[height=60mm,bb=25 169 579    609]{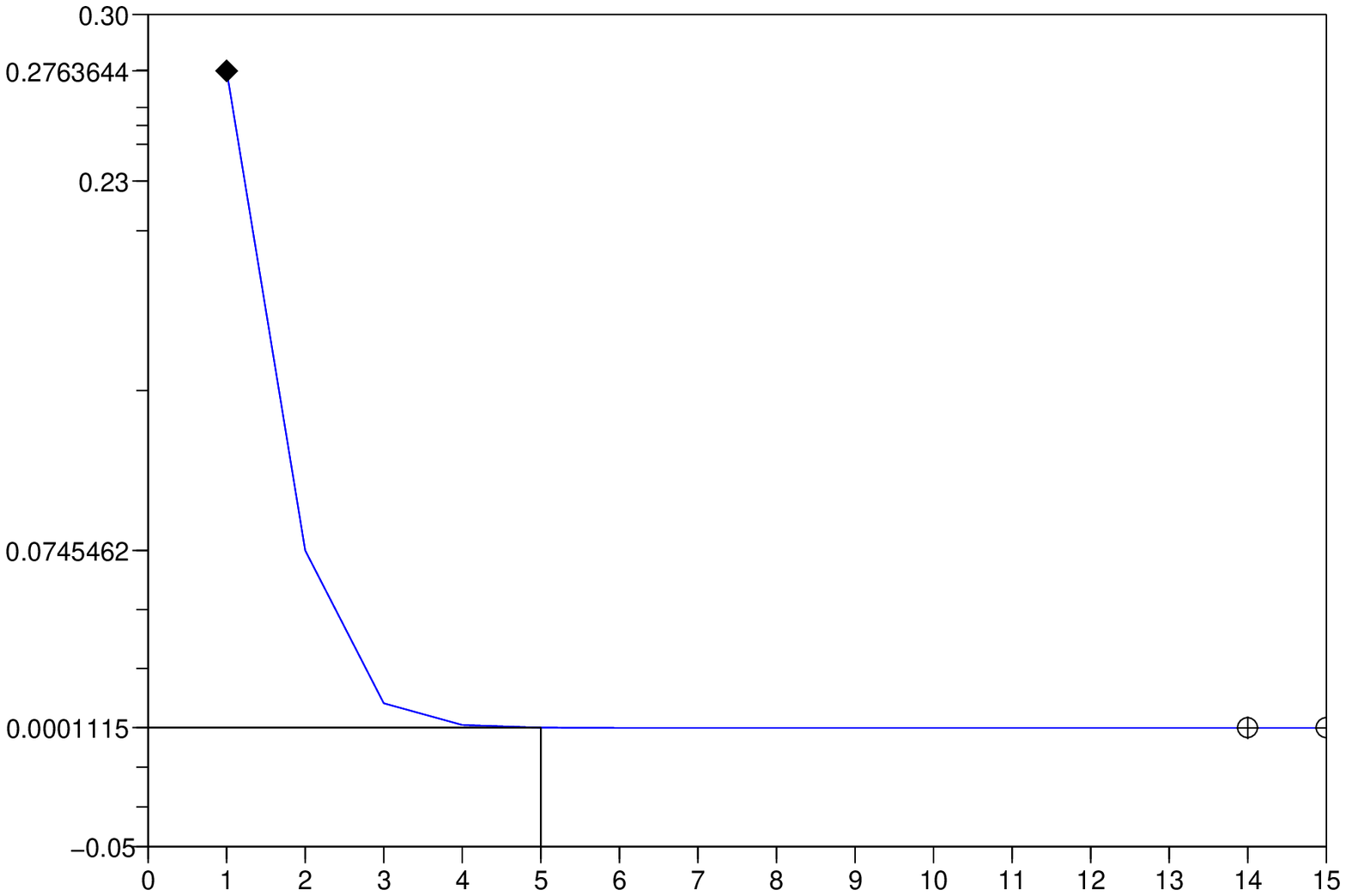}

  \textbf{Figure 2: The accuracy $\delta_\zs{n}$ at each iteration}
\end{center}

 We observe the values of the accuracy: $ \delta_5 \simeq 0.0001$, $\delta_8 \simeq 10^{-08}$ and $\delta_{14}\simeq 10^{-16}$. Moreover
  $\delta_\zs{n}=0$  $\forall n \ge 15$, i.e. numerically
 the limit function $h$ is reached at the $14^{th}$ iteration.

\bigskip

\renewcommand{\theequation}{A.\arabic{equation}}
\renewcommand{\thetheorem}{A.\arabic{theorem}}
\renewcommand{\thesubsection}{A.\arabic{subsection}}
\section{Appendix}\label{sec:A}
\setcounter{equation}{0}
\setcounter{theorem}{0}

\subsection{Properties of the space $(\cX,\varrho_\zs{*})$}

\begin{proposition}\label{Pr.sec:PrL.1}
 $(\cX,\varrho_\zs{*})$ is a complete metrical space.
\end{proposition}
\proof
Indeed, let $(f_\zs{n})_\zs{n\ge 1}$ be a fundamental sequence from
$(\cX,\varrho_\zs{*})$, i.e.
$$
\lim_\zs{m,n\to\infty}
\,\varrho_\zs{*}(f_\zs{n},f_\zs{m})=0\,.
$$
Taking into account that the norm $\|\cdot\|_\zs{*}$ defined in 
\eqref{sec:Mr.3} is equivalent to the norm \eqref{sec:Mr.1-1},
we obtain that this sequence is fundamental in Banach space $\C^{0,1}(\cK)$.
Therefore,  there exists a function
$f\in\C^{0,1}(\cK)$ such that
$$
\lim_\zs{n\to\infty}
\,\varrho_\zs{*}(f_\zs{n},f)=0\,.
$$
The definition of the metrics $\varrho_\zs{*}$ in \eqref{sec:Mr.3} implies immediately that
$f\in\cX$. Hence Proposition~\ref{Pr.sec:PrL.1}. \endproof

\bigskip
\subsection{Properties of the function $\cH_\zs{f}$}

In this subsection we study the smoothness of $\cH_\zs{f}$ with respect to $y$.

\begin{lemma}\label{Le.sec:A.1}
Assume that the condition $\A_\zs{3})$ holds.
Then, for any $0<t\le s\le T$ 
\begin{equation}
\label{sec:A.7}
\max_\zs{1\le i\le m}
\sup_\zs{y\in\bbr^{m}}
\sup_\zs{f\in \cX}
\left|
\frac{\partial }{\partial y_\zs{i}}
\,\cH_\zs{f}(t,s,y)\,
\right|
\le H^{*}\left(
\frac{1}{\sqrt{s-t}}
+1
\right)
\,,
\end{equation}
where $H^{*}$ is given in \eqref{sec:Fr.2}.
\end{lemma}
\proof
First, for any $y\in\bbr^{m}$ we introduce the auxiliary 
Brownian motion on the interval 
$[t,s]$ as
$$
 \xi_\zs{v}=y +\beta (U_\zs{v}-U_\zs{t})\,.
$$

\noindent By making use of this process and the
Girsanov  theorem (see \cite{LiptserShiryaev1977} p. 254)
we can represent the mapping $\cH_\zs{f}$ as
$$
\cH_\zs{f}(t,s,y)=\E\,
f_\zs{1}(s,\xi_{s})
\,e^{\Phi(\xi)}\,,
$$
where $f_\zs{1}=f^{1-q_\zs{*}}$,
\begin{equation}\label{sec:A.8}
  \Phi(\xi) = \int^{s}_\zs{t}
\,Q_\zs{1}(v,\xi_\zs{v})
\,\d v
+\beta^{-2}
\,
\sum^{m}_\zs{l=1}
\int^{s}_\zs{t}
\,
[\alpha(v,\xi_\zs{v})]_\zs{l}\,
\d [\xi_\zs{v}]_\zs{l}
\end{equation}
and
$$
Q_\zs{1}(v,y)=Q(v,y)-
\frac{|\alpha(v,y)|^{2}}{2\beta^{2}}\,.
$$

\noindent
Now we fixe some $1\le i\le m$ and we set
\begin{equation}\label{sec:A.9}
\varsigma_\zs{v}=[\xi_\zs{v}]_\zs{i}=
y_\zs{i}
+\beta\,
\left( [\U_\zs{v}]_\zs{i}
-
[\U_\zs{t}]_\zs{i}
\right)
\,.
\end{equation}
We recall, that  $[x]_\zs{i}$ is the $i$th component of the vector $x\in\bbr^{m}$.
\noindent
Taking the conditional expectation with respect to
$\varsigma_\zs{s}$ we represent $\cH_\zs{f}$ as
\begin{equation}\label{sec:A.10}
\cH_\zs{f}(t,s,y)
=
\int_\zs{\bbr}\,\wh{\cH}_\zs{f}(s,y,z)\,
\p(z,y_\zs{i})
\d z\,,
\end{equation}
where $\wh{\cH}_\zs{f}(s,y,z)=\E\left(
f_\zs{1}(s,\xi_{s})
\,e^{\Phi(\xi)}
|\varsigma_\zs{s}=z
\right)$,
$$
\p(z,y_\zs{i})=\frac{1}{\nu\sqrt{2\pi}}\,
e^{-\frac{(z-y_\zs{i})^{2}}{2\nu^{2}}}
\quad\mbox{and}\quad
\nu^{2}=\beta^{2}(t-s)
\,.
$$
To calculate this conditional expectation
note, that one can check directly that
for any $t< v_\zs{1}<\ldots <v_\zs{k}< s$ and
for any bounded $\bbr^{k}\to\bbr$ function $G$
\begin{equation}\label{sec:A.11}
\E
\left(
G(\varsigma_\zs{v_\zs{1}},\ldots,\varsigma_\zs{v_\zs{k}})
|\varsigma_\zs{s}=z
\right)=
\E
\,
G(B_\zs{v_\zs{1}},\ldots,B_\zs{v_\zs{k}})
\,,
\end{equation}
where  $B_\zs{v}$ is the well-known 
Brownian Bridge on the interval $[t,s]$ (see, for example, \cite{KaratzasShreve1991}, p. 359), i.e.
\begin{equation}\label{sec:A.11-1}
B_\zs{v}=y_\zs{i}
+
\frac{v-t}{s-t}
\left(z-
y_\zs{i}
\right)
+
\varsigma^{0}_\zs{v}
-
\frac{v-t}{s-t}
\varsigma^{0}_\zs{s}
\end{equation}
and $\varsigma^{0}_\zs{v}=
\beta\,
\left( [\U_\zs{v}]_\zs{i}
-
[\U_\zs{t}]_\zs{i}
\right)$.  Moreover, using the representation \eqref{sec:Mm.9} we can represent the stochastic
integral with respect to the $i$th component
in \eqref{sec:A.8} as
\begin{equation}\label{sec:A.12}
\int^{s}_\zs{t}
\,
[\alpha(v,\xi_\zs{v})]_\zs{i}\,
\d [\xi_\zs{v}]_\zs{i}
=
\int^{s}_\zs{t}
\,
\overline{\a}_\zs{i}(v,\xi_\zs{v})\,
\d \varsigma_\zs{v}
+
\int^{s}_\zs{t}
\,
\a_\zs{i,i}(v,\varsigma_\zs{v})\,
\d \varsigma_\zs{v}\,,
\end{equation}
where 
$$
\overline{\a}_\zs{i}(v,x)=
\sum^{m}_\zs{l=1,l\neq i}\,
\a_\zs{i,l}(v,x_\zs{l})
\quad\mbox{for}\quad x=(x_\zs{1},\ldots,x_\zs{m})'\,.
$$
By putting
$$
A_\zs{i}(v,z,y_\zs{i})=\int^{z}_\zs{y_\zs{i}}\,\a_\zs{i,i}(v,u)\d u
$$
 and using the
Ito formula, 
we can rewrite the last stochastic integral 
in \eqref{sec:A.12}
as
$$
\int^{s}_\zs{t}
\,
\a_\zs{i,i}(v,\varsigma_\zs{v})\,
\d \varsigma_\zs{v}
=A_\zs{i}(s,\varsigma_\zs{s},y_\zs{i})-
\int^{s}_\zs{t}\,
\wt{A}_\zs{i}(v,\varsigma_\zs{v},y_\zs{i})\d v\,,
$$
where 
$$
\wt{A}_\zs{i}(v,z,y_\zs{i})=
\int^{z}_\zs{y_\zs{i}}\,
\frac{\partial }{\partial v}
\,\a_\zs{i,i}(v,u)\,\d u
+
\frac{\beta^{2}}{2}
\frac{\partial }{\partial z}\a_\zs{i,i}(v,z)\,.
$$
\noindent Now for any vector $x=(x_\zs{1},\ldots,x_\zs{m})'$ we set
\begin{equation}\label{sec:A.12-1}
\check{x}^{i}=
\left(x_\zs{1},\ldots,x_\zs{i-1},0,x_\zs{i+1},\ldots,x_\zs{m}
\right)'\,.
\end{equation}
Using these  notations one can represent the function \eqref{sec:A.8}
as
$$
\Phi(\xi)=A_\zs{i}(s,\varsigma_\zs{s},y_\zs{i})+
\Phi_\zs{1,i}(\xi,y_\zs{i})\,,
$$
where
\begin{align*}
\Phi_\zs{1,i}(\xi,y_\zs{i})&=\beta^{-2}
\int^{s}_\zs{t}\,\left(
\check{\alpha}^{i}(v,\xi_\zs{v})\right)'\d \xi_\zs{v}
+\beta^{-2}
\int^{s}_\zs{t}\,
\overline{\a}_\zs{i}(v,\xi_\zs{v})
\d \varsigma_\zs{v}\\[2mm]
&+
\int^{s}_\zs{t}\,
Q_\zs{2,i}(v,\xi_\zs{v},y_\zs{i})
\d v
\end{align*}
and 
$Q_\zs{2,i}(v,x,y_\zs{i})=
Q_\zs{1}(v,x)
-\wt{A}_\zs{i}(v,x_\zs{i},y_\zs{i})$.
\noindent
Therefore, by the property \eqref{sec:A.11} 
\begin{equation}\label{sec:A.13}
\wh{\cH}_\zs{f}(s,y,z)=e^{A_\zs{i}(s,z,y_\zs{i})}\E\left(
f_\zs{1}(s,\wh{\xi}^{i}_{s})
\,e^{\Phi_\zs{1,i}(\wh{\xi}^{i},y_\zs{i})}
\right)\,,
\end{equation}
where 
$\wh{\xi}^{i}_\zs{v}=
\left(
[\xi_\zs{v}]_\zs{1},
\ldots,[\xi_\zs{v}]_\zs{i-1},B_\zs{v},[\xi_\zs{v}]_\zs{i+1}\ldots,
[\xi_\zs{v}]_\zs{m}
\right)^{'}$.
Taking into account that
$
\overline{\a}_\zs{i}(v,\wh{\xi}^{i}_\zs{v})
=\overline{\a}_\zs{i}(v,\xi_\zs{v})$ and 
$$
\int^{s}_\zs{t}
\left(
\check{\alpha}^{i}(v,\wh{\xi}^{i}_\zs{v})\right)'\d \wh{\xi}^{i}_\zs{v}
=
\int^{s}_\zs{t}
\left(
\check{\alpha}^{i}(v,\wh{\xi}^{i}_\zs{v})\right)'\d \xi_\zs{v}\,
$$
we get
\begin{align*}
\Phi_\zs{1,i}(\wh{\xi}^{i},y_\zs{i})&=\beta^{-2}
\int^{s}_\zs{t}\,\left(
\check{\alpha}^{i}(v,\wh{\xi}^{i}_\zs{v})\right)'\d \xi_\zs{v}
+\beta^{-2}
\int^{s}_\zs{t}\,
\overline{\a}_\zs{i}(v,\xi_\zs{v})
\d B_\zs{v}\\[2mm]
&+
\int^{s}_\zs{t}\,
Q_\zs{2,i}(v,\wh{\xi}^{i}_\zs{v},y_\zs{i})
\d v\,.
\end{align*}
\noindent
Through the definition of $B_\zs{v}$ in \eqref{sec:A.11-1}
we rewrite this equality as
\begin{align}\nonumber
\Phi_\zs{1,i}(\wh{\xi}^{i},y_\zs{i})&=
\frac{(z-y_\zs{i}-\varsigma^{0}_\zs{s})}{\beta^{2}(s-t)}
\int^{s}_\zs{t}\,
\overline{\a}_\zs{i}(v,\xi_\zs{v})
\d v
+
\int^{s}_\zs{t}\,
Q_\zs{2,i}(v,\wh{\xi}^{i}_\zs{v},y_\zs{i})
\d v\\[2mm]\label{sec:A.13-1}
&+
\int^{s}_\zs{t}\,
\left(\g^{i}(v,\wh{\xi}^{i}_\zs{v})\right)'\,\d U_\zs{v}
\,,
\end{align}
where the function $\g^{i}(v,x)\in\bbr^{m}$ is defined as
$$
[\g^{i}(v,x)]_\zs{j}=\beta^{-1}
\left\{
\begin{array}{cl}
 [\alpha(v,x)]_\zs{j}&\quad\mbox{for}\quad j\neq i\,;\\[2mm]
\overline{\a}_\zs{i}(v,x)&\quad\mbox{for}\quad j= i\,.
\end{array}
\right.
$$
Due to the condition $\A_\zs{3})$  this function is bounded, i.e.
\begin{equation}\label{sec:A.13-2}
|\g^{i}(v,x)|\le 2\beta^{-1} \alpha_\zs{*}\,.
\end{equation}
\noindent
Let us show now, that for any $q>0$
\begin{equation}\label{sec:A.14}
\E e^{q \Phi_\zs{1,i}(\wh{\xi}^{i},y_\zs{i})}
\le
\phi_\zs{q} e^{ \iota_\zs{q}|z-y_i|}\,,
\end{equation}
where $\phi_\zs{q}$ and $\iota_\zs{q}$ are 
given in \eqref{sec:Fr.1}.
First note, that by the condition $\A_\zs{3})$
$$
|\wt{A}_\zs{i}(v,z,y_\zs{i})|
\le 
\alpha_\zs{*}
\,|z-y_\zs{i}|
+
\alpha_\zs{*}
\,
\beta^{2}/2
$$
and, therefore,
$$
Q_\zs{2,i}(v,x)=
Q_\zs{1}(v,x)
-\wt{A}_\zs{i}(v,z,y_\zs{i})
\le Q_\zs{*}+
\alpha_\zs{*}
\,|z-y_\zs{i}|
+
\alpha_\zs{*}
\,
\beta^{2}/2
\,.
$$
Putting $\zeta^{i}_\zs{v}=\g^{i}(v,\wh{\xi}^{i}_\zs{v})$
and using the coefficients defined in \eqref{sec:Fr.1},
 we  obtain, that for any $0<t<s<T$
\begin{align}\label{sec:A.14-1}
\Phi_\zs{1,i}(\wh{\xi}^{i},y_\zs{i})\,\le\,
\int^{s}_\zs{t}\,
\left(\zeta^{i}_\zs{v}\right)'\,\d U_\zs{v}
+\wt{\alpha}_\zs{1}|\varsigma^{0}_\zs{s}|
+
\wt{\alpha}_\zs{2}
|z-y_\zs{i}|
+
\wt{\alpha}_\zs{3}
\,.
\end{align}

\noindent Obviously, that for each $1\le j\le m$ the process
$([\zeta^{i}_\zs{v}]_\zs{j})_\zs{t\le v\le s}$ is 
adapted to the filtration $(\cB^{j}_\zs{v})_\zs{t\le v\le s}$, where
$$
\cB^{j}_\zs{v}=\sigma\left\{
(
[U_\zs{u}]_\zs{j})_\zs{t\le u\le v}
\  , \ 
(\check{U}^{j}_\zs{u})_\zs{t\le u\le s}
\right\}\,.
$$
Therefore, for any $q>0$ 
the process
$$
\varrho^{j}_\zs{v}=
e^{q\int^{v}_\zs{t}\,[\zeta^{i}_\zs{v}]_\zs{j}\,\d [U_\zs{v}]_\zs{j}
-\frac{q^{2}}{2}\int^{v}_\zs{t}\,[\zeta^{i}_\zs{u}]^{2}_\zs{j}\,\d u }
$$
is martingale, i.e. $\E\,\varrho^{j}_\zs{s}\,=1$.
This and \eqref{sec:A.13-2} yield
$$
 \E\,e^{q\int^{s}_\zs{t}\,[\zeta^{i}_\zs{v}]_\zs{j}\,\d [U_\zs{v}]_\zs{j}}
=
\E\,\varrho^{j}_\zs{s}\,
e^{\frac{q^{2}}{2}\int^{s}_\zs{t}\,[\zeta^{i}_\zs{v}]^{2}_\zs{j}\,\d v}
\le e^{q^{2}\beta^{-2}T\alpha^{2}_\zs{*}}\,.
$$
Now we use  the following multidimensional version of the  H\"older inequality.
For any integrated variables  $(\eta_\zs{j})_\zs{1\le j\le  m}$
$$
\E \prod^{m}_\zs{j=1}\, |\eta_\zs{j}|\le 
\, \prod^{m}_\zs{j=1}\ \left(\E\,|\eta_\zs{j}|^{r_\zs{j}}\right)^{1/r_\zs{j}}\,,
$$
where $(r_\zs{j})_\zs{1\le j\le  m}$ are positive numbers
such, that $\sum^{m}_\zs{j=1}\,r^{-1}_\zs{j}=1$. 
This inequality implies
\begin{align*}
\E\,e^{q\int^{s}_\zs{t}\,\g^{i}(v,\wh{\xi}^{i}_\zs{v})\,\d U_\zs{v}}
&=\E \prod^{m}_\zs{j=1}\,
e^{q\int^{s}_\zs{t}\,[\zeta^{i}_\zs{v}]_\zs{j}\,\d [U_\zs{v}]_\zs{j}}\\[2mm]
&\le  \prod^{m}_\zs{j=1}\,
\left(\E
e^{q m\int^{s}_\zs{t}\,[\zeta^{i}_\zs{v}]_\zs{j}\,\d [U_\zs{v}]_\zs{j}}
\right)^{1/m}\le e^{q^{2}\beta^{-2}m^{2}T\alpha^{2}_\zs{*}}\,.
\end{align*}
Now from \eqref{sec:A.14-1} for any $q>0$ we can estimate 
the expectation in \eqref{sec:A.14} in the following way
\begin{align*}
\E e^{q \Phi_\zs{1,i}(\wh{\xi}^{i})}\,
&\le\,
e^{q\wt{\alpha}_\zs{2}\,|z-y_\zs{i}|+q\wt{\alpha}_\zs{3}}
\,\E\,e^{q\int^{s}_\zs{t}\,
\left(\zeta^{i}_\zs{v}\right)'\,\d U_\zs{v}+
 q\wt{\alpha}_\zs{1}|\varsigma^{0}_\zs{s}|}\\[2mm]
&\le 
\,e^{q\wt{\alpha}_\zs{2}\,|z-y_\zs{i}|+q\wt{\alpha}_\zs{3}}
\,\left(\E\,e^{2q\int^{s}_\zs{t}\,
\left(\zeta^{i}_\zs{v}\right)'\,\d U_\zs{v}}
\right)^{1/2}\,
\left(
\E e^{2q\wt{\alpha}_\zs{1}|\varsigma^{0}_\zs{s}|}
\right)^{1/2}
\,,
\end{align*}
where $\wt{\alpha}_\zs{1}=\alpha_\zs{*}\beta^{-2}$, 
$\wt{\alpha}_\zs{2}=\alpha_\zs{*}\,\left(\beta^{-2}+T
\right)$ and $\wt{\alpha}_\zs{3}=T\left(Q_\zs{*}
+\beta^{2}\alpha_\zs{*}/2
\right)
$.
Taking into account here, that for any $a\in\bbr$
$$
\E e^{a	|\varsigma^{0}_\zs{s}|}\le 2 e^{a^{2}\beta^{2}(s-t)/2}\,,
$$
we come to the upper bound \eqref{sec:A.14}.
Obviously, that in \eqref{sec:A.13} the function
$|A_\zs{i}(s,z)|\le \alpha_\zs{*}|z-y_\zs{i}|$. Therefore,
\begin{equation}\label{sec:A.16}
\sup_\zs{f\in\cX}
\,\wh{\cH}_\zs{f}(s,y,z)
\, \le
\phi_\zs{1} e^{(\iota_\zs{1}+\alpha_\zs{*})|z-y_i|}\,.
\end{equation}
 Moreover, from \eqref{sec:A.13}
we get
\begin{align}\nonumber
\frac{\partial }{\partial y_\zs{i}}
\wh{\cH}_\zs{f}(s,y,z)
&= -\a_\zs{i,i}(s,y_\zs{i})
\wh{\cH}_\zs{f}(s,y,z)\\\label{sec:A.17}
&+e^{A_\zs{i}(s,z,y_\zs{i})}\wh{\cH}^{1}_\zs{f,i}(s,y,z)\,,
\end{align}
where
$$
\wh{\cH}^{1}_\zs{f,i}(s,y,z)=\E\left(
f_\zs{1}(s,\wh{\xi}^{i}_{s})
\,e^{\Phi_\zs{1,i}(\wh{\xi}^{i},y_\zs{i})}
\,\frac{\partial }{\partial y_\zs{i}}\,\Phi_\zs{1,i}(\wh{\xi}^{i},y_\zs{i})
\right)\,.
$$
\noindent
Therefore,
$$
\sup_\zs{f\in\cX}\,
\left|
\wh{\cH}^{1}_\zs{f,i}(s,y,z)
\right|
\le
\sqrt{\E\,
e^{2\Phi_\zs{1,i}(\wh{\xi}^{i})}}
\,
\sqrt{\E
\,\Psi^{2}_\zs{i}(t,s)
}\,,
$$
where 
$\Psi_\zs{i}(t,s,y_\zs{i})=\partial \Phi_\zs{1,i}(\wh{\xi}^{i},y_\zs{i})/\partial y_\zs{i}$.
Taking into account here the bound \eqref{sec:A.14} we obtain
$$
\sup_\zs{f\in\cX}\,
\left|
\wh{\cH}^{1}_\zs{f,i}(s,y,z)
\right|
\le
\,(\phi_\zs{2})^{1/2}\,
 e^{ \frac{\iota_\zs{2}}{2}|z-y_i|}
\,
\sqrt{\E\,\Psi^{2}_\zs{i}(t,s)}\,.
$$
Moreover, through \eqref{sec:A.11-1}
and \eqref{sec:A.13-1} we can calculate directly, that
\begin{align*}
\Psi_\zs{i}(t,s,y_\zs{i})&=\sum^{m}_\zs{j=1\,,\,j\neq i}\,
\int^{s}_\zs{t}\,\frac{\partial \a_\zs{j,i}(v,B_\zs{v})}{\partial z}
\,
\frac{(s-v)}{(s-t)}
\,
\d [U_\zs{v}]_\zs{j}\\[2mm]
&+
\int^{s}_\zs{t}\,
\left(\frac{(s-v)}{(s-t)}\Psi_\zs{1,i}(v,\wh{\xi}^{i}_\zs{v},y_\zs{i})
-
\frac{\overline{\a}_\zs{i}(v,\xi_\zs{v})}{\beta^{2}(s-t)}\,\right)
\d v\,,
\end{align*}
where for any $t\le v\le t$ and any vector 
$x=(x_\zs{1},\ldots,x_\zs{m})'$
\begin{align*}
\Psi_\zs{1,i}(v,x,y_\zs{i})&=\frac{\partial Q(v,x)}{\partial x_\zs{i}}
-\sum^{m}_\zs{j=1}\,[\alpha(v,x)]_\zs{j}
\frac{\partial \a_\zs{j,i}(v,x_\zs{i})}{\partial z}
-\frac{\partial \a_\zs{i,i}(v,x_\zs{i})}{\partial v}\\[2mm]
&-\frac{\beta^{2}}{2}\,
\frac{\partial^{2} \a_\zs{i,i}(v,x_\zs{i})}{\partial z^{2}}
+\frac{\partial \a_\zs{i,i}(v,y_\zs{i})}{\partial v}
\,.
\end{align*}

\noindent From the condition $\A_\zs{3})$ we can estimate this function 
$$
\max_\zs{1\le i\le m}
\sup_\zs{(v,x)\in\cK}\,
|\Psi_\zs{1,i}(v,x)|\le
D_\zs{*}+\alpha_\zs{*}
\left(
\alpha_\zs{*}+1+\beta^{2}/2
\right)
\,.
$$
Therefore,
\begin{align*}
\E\,\Psi^{2}_\zs{i}(t,s)&=\E\,\int^{s}_\zs{t}\,
\sum^{m}_\zs{j=1\,,\,j\neq i}\,
\left(\frac{\partial \a_\zs{j,i}(v,B_\zs{v})}{\partial z}
\right)^{2}
\,
\frac{(s-v)^{2}}{(s-t)^{2}}
\,
\d v\\[2mm]
&+
\E\left(
\int^{s}_\zs{t}\,
\left(\frac{(s-v)}{(s-t)}\Psi_\zs{1,i}(v,\wh{\xi}^{i}_\zs{v})
-
\frac{\overline{\a}_\zs{i}(v,\xi_\zs{v})}{\beta^{2}(s-t)}\,\right)
\d v
\right)^{2}
\,.
\end{align*}
This implies directly
$$
\E\,\Psi^{2}_\zs{i}(t,s)\le 
\Psi^{2}_\zs{*}\,,
$$
where the upper bound $\Psi_\zs{*}$ is given in \eqref{sec:Fr.2}.
Therefore,
$$
\sup_\zs{f\in\cX}\,
\left|
\wh{\cH}^{1}_\zs{f,i}(s,y,z)
\right|
\le
\,
\Psi_\zs{*}\,(\phi_\zs{2})^{1/2}
\, e^{ \frac{\iota_\zs{2}}{2}|z-y_i|}
\,.
$$
Now from \eqref{sec:A.16} and \eqref{sec:A.17}
it follows that
$$
\sup_\zs{f\in\cX}
\left|\frac{\partial }{\partial y_\zs{i}}\wh{\cH}_\zs{f}(s,y,z)
\right|
\, \le
\wh{H}_\zs{*}
\, e^{\iota_\zs{*}|z-y_i|}\,,
$$
where 
$\wh{H}_\zs{*}=
\alpha_\zs{*}\,\phi_\zs{1}
+
\Psi_\zs{*}\,(\phi_\zs{2})^{1/2}$ and 
$
\iota_\zs{*}=\max\left(\iota_\zs{1}
\,,\,
\iota_\zs{2}/2
\right)
+\alpha_\zs{*}
$
.
 Now from \eqref{sec:A.10} we obtain
\begin{align*}
\frac{\partial \cH_\zs{f}(t,s,y)}{\partial y_\zs{i}}
&=
\int_\zs{\bbr}\,\frac{\partial\wh{\cH}_\zs{f}(s,y,z)}{\partial y_\zs{i}}\,
\p(z,y_\zs{i})
\d z
\\[2mm]
&
+\int_\zs{\bbr}\,\wh{\cH}_\zs{f}(s,y,z)\,
\frac{(z-y_\zs{i})}{\nu^{2}}\p(z,y_\zs{i})
\d z
\,.
\end{align*}
\noindent 
Taking into account here, that for any $a>0$
$$
\int_\zs{\bbr}\,e^{a |z-y_i|}\,
\p(z,y_\zs{i})
\d z\le
2 e^{\frac{a^{2}\nu^{2}}{2}}
$$
and
$$
\int_\zs{\bbr}\,
\frac{|z-y_\zs{i}|\,e^{a|z-y_i|}}{\nu^{2}}\p(z,y_\zs{i})
\d z
\le 
\,
2a\,e^{\frac{a^{2}\nu^{2}}{2}}
+
\frac{\sqrt{2}}{\nu\sqrt{\pi}}
\,,
$$
we obtain the  upper bound  \eqref{sec:A.7}. 
Hence Lemma~\ref{Le.sec:A.1}.
\endproof

\medskip
\subsection{Properties of the process \eqref{sec:Mr.4}}

In this subsection we study the properties
 of the process $\eta =(\eta^{t,y}_\zs{s})_\zs{t\le s\le T}$

\begin{lemma}\label{Le.sec:A.2}
Under the conditions $\A_\zs{1}$)--$\A_\zs{2}$)
 the process $(\eta^{t,y}_\zs{s})_\zs{t\le s\le T}$ is almost sure
continuously differentiable
 with respect to
$y\in\bbr^{m}$ for any $t\le s\le T$, i.e. for any $1\le i\le m$ there exists almost sure
 the  derivative $\upsilon_\zs{i}(s)=\partial \, \eta_\zs{s}^{t,y} / \partial \, y_\zs{i}$
such that
$$
\sup_\zs{0\le s\le T}
\sup_\zs{y\in\bbr^{m}}
\,
\max_\zs{1\le i\le m}
\left|
\upsilon_\zs{i}(s)
\right|
\le e^{\alpha_\zs{*}T}
\quad\mbox{a.s..}
$$
\end{lemma}
\proof
First we introduce the matrix of the first partial derivatives of the function
$\alpha(v,y$) as
$$
\alpha_\zs{0}(t,z)=\left(
\frac{\partial [\alpha(t,z)]_\zs{k}}{\partial z_\zs{l}}
\,,\quad 1\le k,i\le m
\right)\,.
$$
One can check directly that the processes
$\upsilon_\zs{i}(s)$ satisfies the following
differential equations
$$
\frac{\d}{\d s}\, \upsilon_\zs{i}(s)=A_\zs{s}\,\upsilon_\zs{i}(s)\,,
\quad \upsilon_\zs{i}(t)=e_\zs{i}\,,
$$
where
$A_\zs{s}=\alpha_\zs{0}(s,\eta^{t,y}_\zs{s})$,
$e_\zs{i}=(0,\ldots,0,1,0,,\ldots,0)'$ (only $i$th component is equal to $1$).
Now by applying here the Gronwall-Bellman inequality
we obtain the upper bounds for the derivatives $\upsilon_\zs{i}(s)$. 
Hence Lemma~\ref{Le.sec:A.2}.
\endproof

\medskip
\medskip
\subsection{Properties of the function $\cG$}

Now we study the partial derivatives of the function $\cG(t,s,y)$ defined in
\eqref{sec:Mr.5}. To this end we need the following general result.

\begin{lemma}\label{Le.sec:A.3}
Let $F=F(y,\omega)$ be a $\bbr\times \Omega\to\bbr$ random bounded
function
such that for some nonrandom constant $c^{*}$
$$
\left|
\frac{\d }{\d y}\,F(y,\omega)
\right|\,\le\,
c^{*}
\quad\mbox{a.s.}
\,.
$$
Then
$$
\frac{\d }{\d y}\,\E\,F(y,\omega)=
\E\,\frac{\d }{\d y}\, F(y,\omega)
\,.
$$
\end{lemma}
\noindent This Lemma follows immediately from the Lebesgue dominated convergence theorem.

\begin{lemma}\label{Le.sec:A.4}
Under the conditions $\A_\zs{1}$)--$\A_\zs{2}$) 
there exist the 
 partial derivatives
$(\partial \cG(t,s,y)/\partial y_\zs{i})_\zs{1\le i\le m}$ such that
\begin{equation}\label{sec:A.18}
\max_\zs{1\le i\le m}\,
\sup_\zs{y\in\bbr^{m}}\,
\left|
\frac{\partial \cG(t,s,y)}{\partial y_\zs{i}}
\right|\,\le\,D_\zs{*}\,T\,
e^{(\alpha_\zs{*}+Q_\zs{*})T}
\end{equation}
and
$$
\frac{\partial }{\partial y_\zs{i}}\,\E\,\cG(t,s,y)=
\E\,\frac{\partial }{\partial y_\zs{i}}\cG(t,s,y)
\,.
$$
\end{lemma}
\proof
Lemma~\ref{Le.sec:A.2} implies immediately, that 
$$
\frac{\partial \cG(t,s,y)}{\partial y_\zs{i}}=\cG(t,s,y)\G_\zs{i}(t,s,y)
$$
where $\G_\zs{i}(t,s,y)=
\int^{s}_\zs{t}
\left(Q_\zs{0}(u,\eta^{t,y}_\zs{u})\right)'\upsilon_\zs{i}(u)
\d u$, $Q_\zs{0}(u,z)=\D_\zs{z} Q(u,z)$ and
$\upsilon_\zs{i}(u)=\partial \, \eta_\zs{u}^{t,y} / \partial \, y_\zs{i}$.
Now 
Lemma~\ref{Le.sec:A.2} and Lemma~\ref{Le.sec:A.3} imply directly this lemma. 

\endproof

\medskip
\medskip

\bibliographystyle{plain}
\bibliography{Biblio_Finance}

\end{document}